\documentclass[pre,superscriptaddress,showpacs,preprint]{revtex4-1}
\usepackage{amsmath,graphicx,amssymb,amsfonts}

\begin{document} \title{Granular Solid Hydrodynamics  (GSH):\\ from Quasi-Static Motion to Rapid Dense Flow}
\author{Yimin Jiang} 
\affiliation{Central South University, Changsha 410083, China}
\author{Mario Liu}
\affiliation{Theoretische Physik, Universit\"{a}t T\"{u}bingen,72076
T\"{u}bingen, Germany}

\date{\today}

\begin{abstract} {\sc gsh} is a continuum mechanical theory constructed to qualitatively account for a broad range of granular phenomena. To probe and demonstrate  its width, simple solutions of  {\sc gsh} are related to granular phenomena and constitutive models, including 
(i)~for vanishing shear rates: static stress distribution and propagation of elastic waves; 
(ii)~at slow rates:  critical state, shear band, the models of  hypoplasticity and barodesy; 
(iii)~at higher rates: the MIDI-model, 
rapid dense flow in the Bagnold regime. 
A unified, densely correlated understanding of granular physics 
emerges as a result of these phenomena 
ordered and explained employing a single framework.
\end{abstract}

\maketitle 
\tableofcontents

\section{Introduction\label{intro}} 
\subsection{Engineering Models}
Being a subject of practical importance, elasto-plastic deformation of dense granular media has been under the focus of engineering research for many decades if not
centuries~\cite{schofield,nedderman,wood1990,kolymbas1,kolymbas2,gudehus2010}. The state of 
engineering theories, however, is confusing for physicists: Innumerable
continuum mechanical models compete, employing strikingly different
expressions. In a recent
book on soil mechanics, phrases such as {\em morass of equations} and {\em
jungle of data} were employed as metaphors~\cite{gudehus2010}. 
Moreover, this competition is among theories applicable only to the slow shear rates of elasto-plastic 
deformation, while rapid dense flow (such as heap flow, mud slide or
avalanches) is taken to obey yet rather different equations~\cite{hutter2007}. As a result, 
most engineers believe it is illusory to look for a unified theory capable of simultaneously 
accounting for granular phenomena of arbitrary shear rates: static stress distribution 
at vanishing rates, the rate-independent, elasto-plastic motion at slow 
rates, and  the rapid dense flow at high rates. 

On the other hand, although many theories achieve considerable realism when confined to the effects they were constructed for, they are in essence clever renditions of complex data, not a reflection of the underlying physics. We therefore thought it may be 
worthwhile to try out a new starting point, by focusing on the physics, while leaving  the rich and subtle granular phenomenology aside while constructing 
the theory. The hope was to arrive at a theory that, though not necessarily detailed or realistic in every aspect, is widely applicable over the compete range of shear rates, firmly based in physics, and affords a well founded, transparent understanding. 
A tried and true method to achieve this aim is to construct the so-called {\em hydrodynamic theory} that physicists take to mean the long-wave-length, continuum theory of condensed systems -- in contrast to engineers, who use the word as a synonym for the Navier-Stokes equations.

\subsection{GSH -- A Hydrodynamic Theory\label{intro-1}} The hydrodynamic formalism  was pioneered by Landau~\cite{LL6} and Khalatnikov~\cite{Khal} in the context of superfluid helium, and 
introduced to complex fluids by de Gennes~\cite{deGennes}. Its two crucial points are: 
First, the input in physics that serves mainly to specify the complete set of state variables; and second, the simultaneous consideration of energy and momentum conservation, complete with their respective fluxes. Combining this with thermodynamic considerations, one finds many more constraints, and far less liberty, than the 
engineering approach to constitutive relations, when constructing expressions for the stress. 
This is an advantage especially for granular media: If a choice of random phenomena is rendered correctly by a hydrodynamic theory, chances are that the rest is also adequately accounted for -- because the theory complies with all the mentioned  general principles and is founded primarily on insights into the basic physics of a system, not a subset of experimental data.

Hydrodynamic theories~\cite{hydro-1,hydro-2} have been derived for many condensed 
systems, including liquid crystals
~\cite{liqCryst-1,liqCryst-2,liqCryst-3,liqCryst-4,liqCryst-5,liqCryst-6,liqCryst-7},
superfluid $^3$He~\cite{he3-1,he3-2,he3-3,he3-4,he3-5,he3-6},
superconductors~\cite{SC-1,SC-2,SC-3}, macroscopic
electro-magnetism~\cite{hymax-1,hymax-2,hymax-3,hymax-4},
ferrofluids~\cite{FF-1,FF-2,FF-3,FF-4,FF-5,FF-6,FF-7,FF-8,FF-9}, and
polymers~\cite{polymer-1,polymer-2,polymer-3,polymer-4}.
We contend that constructing a hydrodynamic theory is both useful and possible for granular media: Useful, because it should help to illuminate and order their
complex behavior; possible, because a significant portion has 
already been done. We call it {\sc gsh}, for ``granular hydrodynamic theory.''
It divides granular behavior into three regimes, with the jiggling of the grains -- quantified as the granular temperature $T_g$ -- serving as a switch:
\begin{itemize}
\item At vanishing shear rates, grains hardly jiggle, and $T_g\to0$. Static stress distribution and the propagation of elastic waves are phenomena of this regime. We call it {\em quasi-elastic} because the stress stems from deformed grains and is elastic in origin. 
\item At slow rates, the jiggling increases and $T_g$ is slightly elevated. Although the 
stress is still predominantly elastic, it may now relax: When the grains loose contact with one another briefly, both granular deformation and the associated stress will decrease. This is the {\em hypoplastic regime} where the hypoplastic model~\cite{kolymbas1} and other rate-independent constitutive relations are valid. Typical phenomena are the {\em critical state}~\cite{schofield} and {\em incremental nonlinearity}, or the strikingly different loading and unloading curves. 
\item At high shear rates, we have the rapid dense flow behavior covered by the MIDI 
model~\cite{midi} and Bagnold flow. The jiggling is so strong that it exerts a pressure, and 
viscosities are important. They compete with the elastic stress, becoming dominating  at very high rates.  
\end{itemize}•

\subsection{Validity of General Principles\label{intro-2}} 

There are many arguments in the
literature contending that granular media, being unique, violate general
principles, hence do not have a hydrodynamic theory  (as first conjectured by Kadanoff~\cite{kadanoff}). At closer scrutiny, none of these arguments is watertight. Four of which (rendered in {\em italic}) are as briefly refuted here as is appropriate for an introduction.
\begin{enumerate} \item {\em ``The energy is not conserved in
granular media.''} Although the kinetic energy of the grains is not
conserved, the total energy is, which includes the heat in the grains.

\item {\em ``Fluctuation-dissipation theorem is not valid in granular media.''}
There are two versions of it, one in terms of the granular
temperature $T_g$, the other in terms of the true temperature $T$. The
latter is a general principle. It always holds and is equally applicable to
a block of copper and a pile of sand, quantifying how much, eg., the volume
of each fluctuates. The former is an imperfect analogy, not a general
principle.

\item {\em ``The Onsager Relation does not hold in sand, because the
underlying microscopic dynamics, 
inelastic scattering, is irreversible.''} The true microscopic dynamics in sand is, as
everywhere else, the reversible Schr\"odinger equation for the constituent
atoms.
\item {\em A sand pile has much more gravitational energy than a monolayer 
of grains. Only the latter is in equilibrium, the minimal energy state. The former, being ``jammed'' and prevented to reach the former, is too far off equilibrium for thermodynamics to hold.} Similar to two chambers of air separated by a stuck piston,  a pile of sand at rest is in fact in equilibrium. The air is in equilibrium because all its many degrees of freedom are except one: the position of the piston that upholds a constraint on the volume of the two subsystems.  In a macroscopic body, all elastic degrees of freedom are in equilibrium if the force balance holds, implying the sum of gravitational and elastic energy is minimal. Two elastic bodies, one on top of another, are also in equilibrium if the sum of their energy is minimal -- though there is the constraint that the upper body must not slide with respect to the lower one. A sand pile is many little elastic bodies on top of one another. If they are constrained to stay put, and their total energy is minimal, the pile is in equilibrium.
\end{enumerate}

\subsection{Two-Stage Irreversibility\label{intro-3}} 
\begin{figure}[tbh] \begin{center}
\includegraphics[scale=0.25]{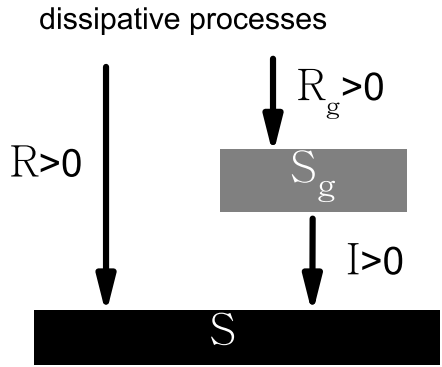}
\end{center}
\caption{\label{2stageIrr}
{\em Two-stage irreversibility}. Dissipative processeses produce either granular entropy $S_g$, or directly thermal entropy $S$. Eventually, $S_g$ is also converted to $S$.}
\end{figure}

To derive the
hydrodynamic theory for granular media, one needs the input of what the
essence of granular physics is. Our working hypothesis is that it is
encapsulated by two notions: {\em two-stage irreversibility} and {\em variable transient elasticity}. The first is related to the three spatial scales of any
granular media: (a)~the macroscopic, (b)~the mesoscopic, intergranular, and (c)~the
microscopic, inner granular. Dividing all degrees of freedom into these three
categories, we treat those of (a) differently from (b,c). Macroscopic
degrees of freedom, such as the slowly varying stress or flow fields, are
specified and employed as explicit state variables, but intergranular and
inner granular degrees are treated summarily: Instead of being specified,
only their contribution to the energy is considered and taken,
respectively, as granular and true heat. So we do not account for the motion of a jiggling grain, only include its strongly fluctuating kinetic and elastic energy as contributions to the granular 
heat, characterized by the granular entropy $s_g$ and temperature $T_g$. Analogously, a phonon, or any elastic vibration within the grain, are taken as part of true heat, part of $s$ and $T$. There are only a handful of macroscopic degrees of freedom (a), innumerable 
intergranular ones (b), and yet many orders of magnitude more inner granular ones (c). So the statistical tendency to equally distribute the energy among all degrees of freedom implies that the energy decays from (a) to (b,c), and from (b) to (c), but never (or hardly ever) backwards. 
This is what we call {\em two-stage irreversibility}, see Fig~\ref{2stageIrr}

The system is in equilibrium only if the true entropy is maximal.
Maximal granular entropy (given in the special cases of the Edward entropy by counting the number of ways to stably build a sand pile, see Sec~\ref{tapping}) would characterize equilibrium, only if there were no energy decay from (a,b) to (c), or when it is slow enough to be neglected. As the ubiquitous inelasticity of granular collisions clearly demonstrate, this is never the case. 

A division into three scales works well when they are clearly
separated, when the system is much larger than the grains. This is indeed a problem with granular media, though one of accuracy, not viability. Scale separation is usually better satisfied in
engineering experiments than in some of physicists. Using glass or steel
beads,  typically larger than sand grains, aggravates the problem.
Same is true of 2D experiments employing less and larger disks. On the
other hand, when there is too little room for spacial averaging, one
can average over time and runs that also get rid of
fluctuations not contained in a continuum theory. 
Moreover, one can go to higher order gradient terms, as we shall do in Sec~\ref{minimal band width}, to capture qualitatively what happens at small length scales -- eg. in shear band or when clogging occurs.

\subsection{Variable Transient Elasticity\label{intro-4}} Our second
notion, {\em variable transient elasticity}, addresses granular plasticity.
The free surface of a granular system at rest is frequently tilted. When
perturbed, when the grains jiggle and $T_g\not=0$, the tilted surface will
decay and become horizontal. The stronger the grains jiggle
and slide, the faster the decay is. We take this as indicative of a system
that is elastic for $T_g=0$, turning transiently elastic for $T_g\not=0$,
with a stress relaxation rate that grows with $T_g$. A relaxing stress is typical of any viscous-elastic system such as polymers. The
unique circumstance here is that the relaxation rate is not a material
constant, but a function of the state variable $T_g$. As we shall see, it
is this dynamically controlled, {\em variable transient elasticity} -- a simple fact at
heart -- that underlies the complex behavior of granular plasticity.
Realizing it yields a most economic way to capture granular rheology.

Employing a strain field rather than the
stress as a state variable usually yields a simpler description, because the former is in
essence a geometric quantity, while the latter contains material parameters such as the stiffness constant. Yet one cannot use the
standard strain field $\epsilon_{ij}$ as a granular state variable, because the relation between stress and $\epsilon_{ij}$ lacks uniqueness when the system is plastic.
Engineering theories frequently divide the strain into two fields, elastic and plastic,
$\epsilon_{ij}=u_{ij}+\epsilon^{(p)}_{ij}$, with the first accounting for
the reversible and second for the irreversible part. They then employ $\epsilon_{ij}$ and $\epsilon^{(p)}_{ij}$ as two independent
strain fields to account for granular plasticity~\cite{Houlsby,Houlsby2}.

We believe that, on the contrary, the elastic strain $u_{ij}$
is the sole state variable, as there is a unique relation between the elastic stress
$\pi_{ij}$ and $u_{ij}$, if the latter is appropriately defined via the elastic energy: Shearing a granular system, a portion of the strain goes into deforming the grains
individually, changing their elastic energy. The rest of the strain is spent sliding
and rolling the grains. Taking $u_{ij}$ as the portion that changes the
energy and deforms the grains, the energy $w$ is by definition a function
of $u_{ij}$ alone. And since an elastic stress $\pi_{ij}$ only exists when the
grains are deformed, it is also a function of $u_{ij}$. Therefore, we
employ $u_{ij}$ as the sole state variable, and discard both
$\epsilon_{ij}$ and $\epsilon^{(p)}_{ij}$. Doing so preserves many useful
features of elasticity, especially the (so-called hyper-elastic) relation,
\begin{equation}\label{1-1} \pi_{ij}=-\partial w(u_{ij})/\partial u_{ij}.
\end{equation} 
This is derived in~\cite{granR2} but easy to understand via
an analogy. Driving up a snowy hill slowly, the car wheels will grip the ground  part of the time, slipping otherwise. (We assume a slowly turning wheel and quickly changing, intermittent stick-slip behavior.) When the wheels do grip, the car moves upward and its gravitational energy $w^{grav}$ is increased. If we divide the wheel's rotation into a gripping (e) and a slipping (p) portion, $\theta=\theta^{(e)}+\theta^{(p)}$, we know we may ignore $\theta^{(p)}$, and compute the torque on the wheel as $\partial 
w^{grav}/\partial\theta^{(e)}$, if the wheel turns sufficiently slowly, same as Eq~(\ref{1-1}). How much the wheel
turns or slips, how large $\theta$ or $\theta^{(p)}$ are, is irrelevant for the
torque.

The only way to find out whether our two hypotheses are appropriate and complete, is to derived the theory and compare its ramifications with experiments. The theory has already been derived, see~\cite{granR2,granR3}, and is called {\sc gsh}. It is briefly repeated and presented in the next chapter. The second step, finding the ramifications of 
{\sc gsh} is a more lengthy process, in the midst of which we are. And this manuscript is an overview on the work done and planned. As mentioned, {\sc gsh} has three rate 
regimes, given by: 
\begin{itemize}
\item The {\em quasi-elastic regime} of vanishing shear rates, possibly below $10^{-5}$ s$^{-1}$, with a quadratically vanishing granular temperature, $T_g\sim\dot\gamma^2\to0$. The stress $\sigma_{ij}=\pi_{ij}$  is purely elastic, as given by Eq~(\ref{1-1}).  

\item The {\em hypoplastic regime} of low shear rates, possibly between $10^{-3}$ and 1/s, where the engineering theory of hypoplastic model~\cite{kolymbas1} holds. The stress $\sigma_{ij}=(1-\alpha)\pi_{ij}$ is still elastic, but softer 
by a factor typically between 0.2 and 0.3. Granular temperature is more elevated, allowing stress relaxation. With $T_g\sim\dot\gamma$, we have rate-independence.  
This regime is frequently hailed as the quasi-static one, because it seems slow, is rate-independent, and because the even slower quasi-elastic regime is (as we shall see) hard to 
observe. We note that the hypoplastic regime, being characterized by stress relaxation,  is dissipative. It therefore cannot possibly be quasi-static, implying a consecutive visit of neighboring equilibrium states. 

\item The {\em rapid flow regime}, for shear rates well above 1 s$^{-1}$. We still have $T_g\sim\dot\gamma$, but it is no longer small. Therefore, the $T_g$-generated, seismic pressure $P_T\sim T_g^2\sim\dot\gamma^2$ and the viscous shear stress $\sigma_s\sim T_g\dot\gamma\sim\dot\gamma^2$ become significant and compete with the elastic contribution $\pi_{ij}$. This is where the MIDI model and Bagnold flow hold. As both the pressure and the shear stress may be written as $e_1+e_2\dot\gamma^2$, where $e_1$ is the elastic, and $e_2$ the seismic or viscous, contributions,  we have the purely quadratic dependence of the Bagnold flow for $e_2\dot\gamma^2\gg e_1$, and hypoplastic rate-independence for  $e_2\dot\gamma^2\ll e_1$. 
\end{itemize}•

\section{Brief Presentation of GSH \label{GSH}    } 

\subsection{Complete Set of State Variables}
In accordance to our understanding of granular media's basic physics, its state variables are: the granular entropy $s_g$ and the elastic strain $u_{ij}$, in addition to the usual variables: the density $\rho$, the momentum density $\rho v_i$, the true entropy $s$. Denoting the
energy density (in the rest frame, $v_i=0$) as $w=w(\rho, s, s_g, u_{ij})$,
we define the conjugate variables as: 
\begin{equation}\label{2-2} \mu\equiv\frac{\partial w}{\partial\rho},\quad
T\equiv\frac{\partial w}{\partial s},\quad T_g\equiv\frac{\partial
w}{\partial s_g},\quad \pi_{ij}\equiv-\frac{\partial w}{\partial u_{ij}},
\end{equation} 
where $\mu$ is the  chemical potential, $T$ the temperature, $T_g$ the 
granular temperature, and $\pi_{ij}$ the elastic stress. These are given once the energy $w$ is.
 %
Next, in Sec~\ref{equicond}, equilibrium conditions will be
derived, formerly, in terms of the energy and its conjugate variables, whatever $w$ is. Then, in Sec~\ref{granEn}, an example for $w$ will be given, and the conjugate variables calculated -- with the help of which the equilibrium conditions are rendered  explicit.

A complete set of state variables is one that determines a
unique macroscopic state of the system. If a set is given, there is no room
for ambiguity, for ``history-" or ``preparation-dependence." Conversely, any such
dependence indicates that the set is incomplete, see eg. the discussion in Sec~\ref{hdvhv}. In the hydrodynamic
approach, a physical quantity is a state variable if (and only if) the
energy $w$ depends on it. We assume the above set is
complete.

Having specified the thermodynamic energy $w$ as a function of relaxing
variables such as $s_g$, we employ in effect a generalized notion of equilibrium,
and treat a state with a finite $s_g$ as being in {\em quasi-equilibrium}. From a statistical 
mechanical point of view, this is a {\em constrained equilibrium}, because we are considering 
only those micro-states that are compatible with the given 
value of $s_g$. (An example for such a relaxing thermodynamic
variable is the magnitude of an order parameter in a Ginzburg-Landau
theory, say the superfluid density $\rho_s$, cf.~\cite{Khal}.) Such a
variable needs to be macroscopically slow, so microscopic variables have
time to adjust to its value. Since $s_g$ typically varies on the scale of
0.1-1 ms in the dense limit, much slower than any microscopic time scales,
$s_g$ is a valid macroscopic and quasi-thermodynamic variable. [The notion of
quasi-equilibrium also holds as local equilibrium, implying
%
$w[{\bf r},t]=w[\rho({\bf r},t), s({\bf r},t),
s_g({\bf r},t), u_{ij}({\bf r},t)]$. 
%
Analogous equations hold for all the conjugate variables, see Eqs~(\ref{2-2}).] 

\subsection{Statics\label{gsh-sta}} 

\subsubsection{Equilibrium Conditions\label{equicond}} 
Requiring maximal entropy $\int s\,{\rm d}^3r$
with appropriate constraints (such as given energy $\int w\,{\rm d}^3r$ and mass $\int \rho\, {\rm d}^3r$), one 
obtains the equilibrium conditions for the state variables in terms of
their conjugate variables. In granular media, remarkably, this universally
valid procedure leads to two distinct sets of equilibrium conditions, the
solid and the fluid one~\cite{granR2,gudehus-jl,granR3}. Maximizing the entropy, we first
obtain the condition of uniform true temperature $\nabla_iT=0$, and the
requirement that the granular temperature vanishes, $T_g=0$. Usually, $T_g$
vanishes quickly, and if it does, the density is not independent from
the elastic strain, $d\rho/\rho=-du_{\ell\ell}$. They share a
common condition that we identify as the solid one,
\begin{eqnarray}
\label{2a-1}
\nabla_i(\pi_{ij}+P_T\delta_{ij})=\rho\, {\rm g}_i,\,\,
\\
P_T\equiv-\partial(wV)/\partial V
\stackrel{T_g\to0}{\longrightarrow}0,
\end{eqnarray} 
where ${\rm g}_i$ is the gravitational constant, $\pi_{ij}$ the elastic stress,
$P_T$ the usual expression for the fluid pressure, and $V$ the volume.  (The derivative is taken at constant $M=\rho V$.) We may equivalently calculate $P_T$ as 
$-\partial(w/\rho)/\partial(1/\rho)=\rho^2\partial(w/\rho)/\partial\rho$, holding constant $s/\rho, s_g/\rho$. With the energy expression  $w$ of the next Sec~\ref{granEn}, $P_T\sim T_g^2$ is the pressure exerted by jiggling grains. We therefore call it the {\em seismic pressure}~\cite{gudehus-jl}.  Clearly, equilibrium condition Eq~(\ref{2a-1}), expressing force balance, is logically the result of maximal true entropy. 

If $T_g$ is kept finite by external perturbations, the system may further increase its entropy by independently varying $\rho$ and $u_{ij}$, to arrive at the fluid equilibrium. It is characterized by two conditions, the first with respect to $u_{ij}$, and the second with respect to $\rho$: 
\begin{equation}\label{2a-2} \pi_{ij}=0, \quad
\nabla_iP_T=\rho\, {\rm g}_i. \end{equation} 
The first condition requires shear stresses to vanish in equilibrium, and free surfaces to be horizontal. The second condition is that governing reversible compaction, a phenomenon one arrives at after keeping $T_g$ finite (such as by tapping) for a long time, see Sec~\ref{compaction}.

\subsubsection{Granular Energy\label{granEn}} 

Interested in hard grains that are slightly excited, implying small $u_{ij},s_g$, we look for the respective lowest order terms in the energy. (As we are not, at present, interested in thermal effects such as thermal expansion, the energy's dependence on the true entropy is not discussed.) Denoting $\Delta\equiv -u_{\ell\ell}$, $P_\Delta\equiv\pi_{\ell\ell}/3$, $u_s^2\equiv
u^*_{ij}u^*_{ij}$, $\pi_s^2\equiv \pi^*_{ij}\pi^*_{ij}$, where
$u^*_{ij},\pi^*_{ij}$ are the respective traceless tensors, we take the energy
to be 
\begin{eqnarray}\label{2b-0}w&=&w_T+w_\Delta
\\ \label{2b-1}
w_T&=&{s_g^2}/({2\rho b}),\quad T_g={s_g}/({\rho b}), 
\\\label{2b-2}
w_\Delta&=&\sqrt{\Delta }(2 {\mathcal B} \Delta^2/5+ {\mathcal A}u_s^2),
\\\label{2b-2a} 
\pi_{ij}=\sqrt\Delta({\cal B}\Delta&+&{\cal A}
{u_s^2}/{2\Delta})\delta _{ij}-2{\cal A}\sqrt\Delta\, u_{ij}^*, 
\\\label{2b-2b} 
P_\Delta=\sqrt\Delta({\cal B}\Delta&+&{\cal A}
{u_s^2}/{2\Delta}),\, \pi_s=-2{\cal A}\sqrt\Delta\, u_s. 
\end{eqnarray} 
Note $u_{ij}$ and $\pi_{ij}$ are colinear and have the same principal axes. The contribution 
$w_T$ is an expansion in $s_g$. The quadratic term is the lowest order one
because we require $s_g,T_g=0$ to be a minimum of the energy. (This is the same
argument as in a Ginzburg-Landau expansion, though without the fourth order term or the phase transition.) As it will turn out, see Eq~(\ref{2b-5}) below,  this lowest order term is in fact sufficient to account for fast dense flow and the gaseous state. In this sense, $s_g$ and $T_g$ are always small.

Next, we compare $T_g$ to the gaseous  granular temperature $T_G$, defined as  $2/3$ of the kinetic energy per particle, see~\cite{luding2009}. Being a general expression for granular heat, $w_T$  includes the quickly fluctuating part of both the kinetic and elastic energy. But in the dilute limit, when the elastic contribution may be neglected, one can take $w_T$ as equal to $3\rho T_G/2\langle m\rangle$ (with $\langle m\rangle$ the average mass of a grain), and identify
\begin{equation}\label{2b-2c} 3T_G/\langle m\rangle = b\, T_g^2. 
\end{equation} 

Fixing the density-dependence of the coefficient $b$ immediately yields an
expression for the seismic pressure $P_T\equiv-\partial(w/\rho)/\partial(1/\rho)$. [There is also a contribution from $w_\Delta\sim\Delta^{2.5}$, because ${\cal A, B}$ depend on the density, see Eq~(\ref{2b-4}). It is neglected because it is always much
smaller than $P_\Delta\sim\Delta^{1.5}$ for small $\Delta$.] We take 
\begin{equation}\label{2b-5}
b=b_0\left(1-\frac\rho{\rho_{cp}}\right)^a,\quad
P_T=\frac{ \rho^2\,a b\, T_g^2}{2(\rho_{cp}-\rho)}, \end{equation}
with both $b_0$ and $a$ being positive numbers. Given Eq~(\ref{2b-2c}) (noting the density dependence of $b$), this is essentially the familiar pressure expression $\sim
T_G/(\rho_{cp}-\rho)$, see eg.~\cite{Bocquet}.    

The second term $w_\Delta$ of Eq~(\ref{2b-1}), with ${\cal A,B}>0$, is the elastic contribution. Its order of 2.5 is important for many granular features, especially {\em stress-induced anisotropy} (see below) and the {\em convexity transition}, discussed in Sec~\ref{yield surfaces}. The associated stress expression $\pi_{ij}$ has been validated for the following circumstances, achieving good to satisfactory agreement: 
\begin{itemize}
\item Static stress distribution in three classic geometries: silo, sand pile,
point load on a granular sheet, calculated using the equilibrium condition, Eq~(\ref{2a-1}), see~\cite{ge1,ge2}. 
\item Small-amplitude stress-strain relation, see~\cite{kuwano2002,ge3}.
\item Anisotropic propagation of elastic waves, see~\cite{jia2009,ge4}. 
\end{itemize}
An explanation of ``stress-induced anisotropy'': In  linear elasticity $w\sim u_s^2$, we have constant second derivatives $\partial^2w/\partial u_s^2$, and the velocity of a elastic wave $\sim\sqrt{\partial^2w/\partial u_s^2}\,$ does not depend on the strain, or equivalently, the stress. For any exponent other than 2, the velocity depends on the stress, and is anisotropic if the stress is.  

Note that the energy $w=w_T+w_\Delta$ vanishes when the grains are neither deformed nor jiggling: $w\to0$ for $s_g, u_{ij}\to0$. This implies a lack of
interaction among the grains. If there were any, there would be a third term in
$w$ that is a function of $\rho$ alone.

\subsection{Yield Surfaces\label{yield surfaces}} 

In a space spanned by stress components and the density, there is a surface  that 
divides two regions in any granular media, one in which the grains necessarily move, 
another in which they may be at rest. This surface is usually referred to as {\em the 
yield surface}.  Aiming to make its definition more precise, we take the yield surface 
to be the divide between two regions, one in which elastic solutions are stable, and another in which they are not. Clearly, the medium may be at rest for a given stress only if an appropriate elastic solution is stable. 
Since the elastic energy of any solution satisfying the equilibrium condition Eq~(\ref{2a-1}) is an extremum, the energy is convex and minimal in the stable region, concave and maximal in the unstable one ---in which infinitesimal perturbations suffice to destroy the solution.

\subsubsection{The Coulomb Yield Surface\label{Druck-Prager}}
The elastic energy of Eq~(\ref{2b-2}) is convex only for 
\begin{equation}\label{2b-3} u_s/\Delta\le\sqrt{2{\cal B}/{\cal A}} \quad
\text{or}\quad \pi_s/P_\Delta\le\sqrt{2{\cal A}/{\cal B}},
\end{equation} 
turning concave if the condition is violated. The second constraint may be derived by rewriting Eq~(\ref{2b-2b}) as 
\begin{equation}\label{2b-3b} 
\frac{4P_\Delta}{\pi_s}=\frac{2{\cal B}}{{\cal
A}}\frac{\Delta}{u_s}+\frac{u_s}{\Delta}, 
\end{equation} 
which shows
$P_\Delta/\pi_s=\sqrt{{\cal B}/2{\cal A}}$ is minimal for
$u_s/\Delta=\sqrt{2{\cal B}/{\cal A}}$. This corroborates the behavior that no
granular system stays static if the shear stress is too large for given
pressure. 

We take ${\cal B}/{\cal A}$ to be density independent and approximately
\begin{equation}
\label{2b-3a} {\cal B}/{\cal A}\approx5/3. 
\end{equation}
Therefore, we only need to specify the density dependence of $\cal B(\rho)$, taking it as
\begin{eqnarray}\label{2b-4} {\cal B}&=&{\cal B}_0
[(\rho-\bar\rho)/(\rho_{cp}-\rho)]^{0.15},\\\nonumber
\bar\rho&\equiv&(20\rho_{\ell p}-11\rho_{cp})/9, 
\end{eqnarray} 
with ${\cal B}_0>0$ a material constant. This expression accomplishes three
things at once:
\begin{itemize}
\item The energy is concave for any density smaller than the
random loose one $\rho_{\ell p}$, implying no elastic solution exists there. 
\item The energy is convex between the random loose density $\rho_{\ell p}$ and the random close one $\rho_{cp}$, ensuring the stability of any elastic
solutions in this region. In addition, the density dependence of sound
velocities as measured by Harding and Richart~\cite{hardin} is well
rendered by $\sqrt{\cal B}$. 
\item The elastic energy diverges, slowly, at
$\rho_{cp}$, approximating the observation that the system becomes orders of magnitude stiffer there.
\end{itemize}

\subsubsection{Yield Stress versus the Critical State}
A widespread confusion is addressed in this section. The
yield surface of Eq~(\ref{2b-3}) defines a yield shear stress for a given
pressure. Many textbooks identify this stress with the highest stress
achieved in an approach to the {\em critical state}, and draw conclusions
based on this identification. Their justification is that the approach is typically
executed at low enough shear rates to be considered {\em quasi-static}. 
We contend that a true {\em quasi-static motion} is one that visits a series of static, equilibrium states, with $T_g\to0$. This happens, as mentioned above,  only during  {\em quasi-elastic motion}, see also Sec~\ref{3regimes}. The rate-independent,  hypoplastic motion, taking place during an approach to the critical state, produces an elevated $T_g$ and is strongly dissipative.   
Therefore, the instability discussed here and the critical state discussed in Sec~\ref{critical state} are two distinct concepts, static versus dynamic. 

The first is a convexity transition of the elastic energy, to be probed by quasi-elastic motion at vanishing $T_g$. The second is a stationary solution of  the
evolution equation for the elastic strain $u_{ij}$, and is comparable to the stationary solution of any diffusion equation. 
The two shear stresses are frequently similar in magnitude, but the yield stress given by Eq~(\ref{2b-3}) needs to be larger than the highest shear stress achieved during the approach to the critical state, see Fig~\ref{yield-cs-pai} below. Otherwise, the system will abandon the approach and develop shear bands instead, considered in Sec~\ref{sb}.

\subsubsection{The Virgin Consolidation Surface}

\begin{figure}[b] \begin{center}
\includegraphics[scale=0.38]{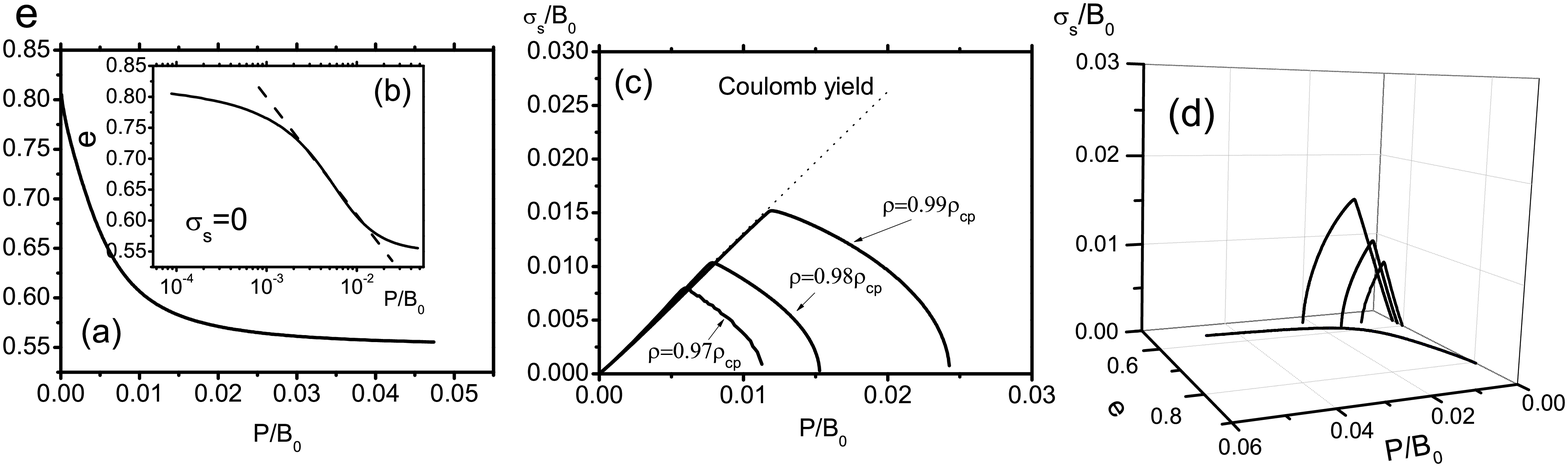}
\end{center}
\caption{\label{fig1}
Granular yield surfaces for $T_g=0$, as a function of the pressure $P$, shear stress $\sigma_s=\pi_s$, and the void ratio $e$, as calculated from the convexity transition of the energy:
$w/{\cal B}_0=\left(\frac{\rho -\rho _{lp}^{\ast}}{\rho _{cp}-\rho}\right)^{0.15}\sqrt{\Delta }\left(\frac25\Delta ^{2}+\frac{1}{\xi}u_{s}^{2}\right)-\left( {\cal D} _{1}\Delta ^{3}+{\cal
D}_{2}u_{s}^{2}\Delta+{\cal D}_3 u_s^4 \right) $, with $\xi\equiv\frac{\cal B}{\cal A} =5/3$, ${\cal D}_{1}=1$, ${\cal D}_{2},{\cal D}_3=2$, and $\rho _{lp}^{\ast }=0.67\rho
_{cp}$ (implying $\rho _{lp}=0.85\rho _{cp}$).
The (a,b) are at $\sigma _{s}=0$, where the inset has a logarithmic scale, while the curves of (c) are at the indicated densities. The dashed straight lines in (b,c) are, respectively, the engineer formula $e=e_{0}-k\ln P$ and the Coulomb yield line. The curves of (d) are the same as in (a,c), though now in the space spanned by $e,P,\sigma_s,$. So this gives the boundary surface of static states.}
\end{figure} 

As depicted in Fig.~\ref{fig1},  granular media possess more yield surfaces in
the space spanned by the pressure $P_\Delta$, shear stress $\sigma_s=\pi_s$, and the void ratio $e$, where $e\equiv1/\phi-1$ ($\phi\equiv\rho/\rho_g$ is the packing fraction, and $\rho_g$  the bulk density of the grains). First,
for given $e$, there is a maximal pressure that a granular system can
sustain before it collapses, implying a yield surface as depicted in (a) of Fig~\ref{fig1}. 
This is a boundary that sand at rest will not cross when compressed. Instead, it will
collapse, becoming more compact, with a smaller $e$, coming to rest at a point close to the curve, never above it. In soild mechanics textbooks, the boundary is frequently referred to as the 
{\em virgin consolidation line}. (The inset, (b) of Fig~\ref{fig1}, has a logarithmic scale. It serves to demonstrate that the standard formula  $e=e_{0}-k\ln P$ from engineering textbooks do not go to $\rho _{lp}$ and $\rho _{cp}$,  for $P\rightarrow 0$ and $\infty $, respectively.)

This qualitative behavior is easily accounted for by higher-order strain terms. Taking ${\cal D}_1,{\cal D}_2, {\cal D}_3>0$ as functions of the density, we choose
\begin{equation}\label{2b-5a} -{\cal B}_0({\cal D}_1\Delta^3
+{\cal D}_2\Delta u_s^2+{\cal D}_3 u_s^4), 
\end{equation} 
to be added to $w_\Delta$,
Eq~(\ref{2b-2}). Consider first $u_s^2=0$. If $\Delta$ is large enough, the
term $-{\cal D}_1\Delta^3$, with a negative second derivative, will
work against ${\cal B}\Delta^{2.5}$ and turn $w_\Delta$ concave. The value of
$\Delta$ at which this happens, call it $\Delta_c$, is given by
$\sqrt{\Delta_c}=5{\cal B}(\rho)/8{\cal D}_1(\rho)$. As ${\cal B}$ diverges
at $\rho_{cp}$, so does $\Delta_c$. If $\Delta_c(\rho)=0$ for
$\rho=\rho_{\ell p}$, ${\cal D}_1(\rho)$ will have to diverge there.

Next consider $u_s^2\not=0$. If ${\cal D}_2,{\cal D}_3=0$, the yield lines in the
space spanned by $P_ \Delta,\pi_s$ for given density would be vertical
lines. The presence of $-{\cal D}_2\Delta u_s^2$ and  $-{\cal D}_3 u_s^4$ reduce the value of
$\Delta$ (or $P_ \Delta$) for growing $u_s$ (or $\pi_s$), bending the lines
to the left. Although qualitative figures of these curves -- frequently
referred to as {\em ``caps"} -- abound in textbooks~\cite{schofield,nedderman}, we did not find enough granular data, especially not a mathematical expression that we could have used to fix
${\cal D}_1,{\cal D}_2,{\cal D}_3$.

\subsubsection{More Yield Surfaces}
The yield surface of Eq~(\ref{2b-3}) is usually referred to as the
Drucker-Prager approximation of the Coulomb yield surface. The actual Coulomb
law is different. And there are more yield laws, such as
{\em Lade-Duncan}~\cite{lade-duncan} or {\em Matsuoka-Nakai}~\cite{matsuoka}.
Engineers choose among them depending on the system, personal preferences
and experiences, apparently without a commonly accepted rule. By including
the third strain invariant $u_t^3\equiv u^*_{ij}u^*_{jk}u^*_{ki}$ into
Eq~(\ref{2b-2}), 
\begin{equation}\label{2b-6} w_\Delta= {\mathcal B}
\sqrt{\Delta }\left(\frac25\Delta^2+ \frac{\mathcal A} {\mathcal B} u_s^2
-\frac{\mathcal C} {\mathcal B} \frac{u^3_t}\Delta\right), \end{equation}
with ${\cal A,B,C}>0$, it is in fact possible to account for many of them
at the same time. (Note the new term is of the lowest order, 2.5.) Depending on how large $\cal C/B$ is, the convexity transition takes place at yield surfaces that are numerically
indistinguishable from the respective yield law. Because a single expression
is employed, and because intermediate yield laws are also possible, this is
a simplifying and unifying step. Evidence that these laws may be unified using Eq~(\ref{2b-6}), and that $u_{ij},\pi_{ij}$ retain their colinearity, is given in~\cite{3inv}.
Defining the friction angle as
$\varphi\equiv\arcsin\sqrt{3/(6P_\Delta^2/\pi_s^2-1)}$, the Coulomb,
Drucker-Prager, and Lade-Duncan yield laws are given respectively as
\begin{eqnarray} \frac{\pi_3-\pi_1}{\pi_3+\pi_1}&=&\sin\varphi,
\quad
\frac{\pi_s}{P_\Delta}=\frac{\sqrt{6}\sin\varphi}{\sqrt{3+\sin^2\varphi}},
\\\nonumber \frac{\pi_1\pi_2\pi_3}{P^3_\Delta}&=&\frac{27
(1-\sin\varphi)\cos^2\varphi}{(3-\sin\varphi)^3}. \end{eqnarray}

\subsection{Dynamics \label{dynamics}} 
\subsubsection{Structure of the Dynamics}

Next, we specify the evolution equations for the state variables. The equation for the elastic strain, assuming both $u_{ij},v_{ij}$ are uniform, is~\cite{granR2}
\begin{equation}\label{2c-6a}
\partial_tu_{ij}-v_{ij}+\alpha_{ijk\ell}v_{k\ell}=-(\lambda_{ijk\ell}T_g)\,u_{k\ell}
\end{equation} 
(where  $v_{ij}\equiv\frac12(\nabla_iv_j+\nabla_jv_i)$ is the shear rate,
$v^*_{ij}$ its traceless part, and $v_s^2\equiv v^*_{ij}v^*_{ij}$).  
If $T_g$ is finite, grains jiggle and briefly
lose contact with one another, during which their deformation is
partially lost. Macroscopically, this shows up as a relaxation of
$u_{ij}$, with a rate that grows with $T_g$, and vanishes for
$T_g=0$. So the lowest order term in an $T_g$-expansion is
$\lambda_{ijk\ell}T_g$. With the elastic energy a convex function, the (negative) elastic stress $-\pi_{ij}=\partial w/\partial u_{ij}$ is a monotonically increasing function of $u_{ij}$. Therefore,  $-\pi_{ij}, u_{ij}$
decrease at the same time. And Eq~(\ref{2c-6a}) accounts for the stress relaxation discussed in the introduction. 

The Onsager coefficient $\alpha_{ijk\ell}$ is an off-diagonal element. Dividing $u_{ij}$ into $\Delta\equiv-u_{\ell\ell}$, $u_{ij}^*$, and specifying the matrices $\alpha_{ijk\ell},\lambda_{ijk\ell}$ with two elements each, Eq~(\ref{2c-6a}) is written as
\begin{eqnarray}
\label{2c-7xx}
\partial_t\Delta+(1-\alpha )v_{\ell\ell} -\alpha_1u^*_{ij}v^*_{ij}
=-\lambda_1T_g\Delta, 
\\\label{2c-8} 
\partial_tu^*_{ij}-(1-\alpha )v^*_{ij}
= -\lambda T_gu^*_{ij},
\\\label{2c-9xx}
\partial_tu_s-(1-\alpha )v_s= -\lambda T_gu_s.
\end{eqnarray} 
The third equation is valid only if strain and rate are colinear, $u^*_{ij}/|u_s|=v^*_{ij}/|v_s|$. This is frequently the case, because any 
component of $u_{ij}$ not co-linear with $v_{ij}$ relaxes to zero quickly. 
The coefficients  $\alpha$ (assuming  $0<\alpha <1$) describes a softening, or more precisely a reduced gear ratio: The same shear rate yields a smaller deformation, $\partial_tu_{ij}=(1-\alpha)v_{ij}\cdots$, but 
acts also at a smaller stress, $\sigma_{ij}=(1-\alpha)\pi_{ij}\cdots$.  $\alpha_1$ accounts for the fact that shearing granular media will change the compression $\Delta$, implying, as we shall see, {\em dilatancy} and {\em contractancy}. (More Onsager coefficients are permitted by symmetry, but  excluded here to keep the equations as simple as possible.) 

Next are the continuity equations for density and momentum
density, 
\begin{equation}\label{2c-1} \partial_t\rho+\nabla_i(\rho
v_i)=0,\,\, \partial_t(\rho v_i)+\nabla_j(\sigma_{ij} +\rho v_iv_j)=0,
\end{equation} 
where the stress tensor $\sigma_{ij}=P\delta_{ij}+\sigma^*_{ij}$ (with $\sigma^*_{ij}$ the traceless part) is determined by general principles~\cite{granR2,granR3}  as
\begin{eqnarray} \label{2c-2}
P\equiv\sigma_{\ell\ell}/3=(1-\alpha )P_\Delta+P_T-\zeta_gv_{\ell\ell},
\\\label{2c-2a}
\sigma^*_{ij}=(1-\alpha)\pi_{ij}^*-\alpha_1u^*_{ij}P_\Delta -\eta_gv^*_{ij},
\\ \label{2c-2b}
\sigma_{s}=(1-\alpha )\pi_s-\alpha_1u_sP_\Delta
-\eta_gv_s. 
\end{eqnarray} 
Again, the third equation (with $\sigma_s^2\equiv\sigma_{ij}^*\sigma_{ij}^*$) is
valid only if $\pi_{ij}^*$ and $v^*_{ij}$ are colinear,
$\pi_{ij}^*/|\pi_s|=v^*_{ij}/|v_s|$. 
The pressure $P$ and shear stress $\sigma_s$ contain elastic contributions
$\sim\pi_s,P_\Delta$ from Eq~(\ref{2b-2b}), the seismic pressure $P_T\sim T_g^2$ from Eq~(\ref{2b-5}), and viscous contributions $\sim\eta_g,\zeta_g$. The  off-diagonal Onsager coefficients $\alpha ,\alpha_1$ (introduced in the equation for the elastic strain $u_{ij}$) soften and mix the elastic stress components. The term preceded by $\alpha_1$ is smaller by an order in the elastic strain, and may frequently be neglected.

The balance equation for granular entropy $s_g=b\rho T_g$ is 
\begin{eqnarray}\label{2c-4} 
\partial_ts_g+\nabla_i(s_gv_i-\kappa\nabla_iT_g)=
\\\nonumber
(\eta_g v_s^2+\zeta_gv^2_{\ell\ell}-\gamma T_g^2)/T_g. 
\end{eqnarray}
Here, $s_gv_i$ is the convective, and $-\kappa\nabla_iT_g$ the diffusive flux. $\eta_g v_s^2$ accounts for viscous heating, for the increase of $T_g$ because 
macroscopic shear rates jiggle the grains. A compressional rate
$\zeta_gv^2_{\ell\ell}$ does the same, though not as efficiently~\cite{granL3}. The term $-\gamma T_g^2$ accounts for the relaxation of $T_g$, ie., for the conversion of intergranular energy into inner granular one.

\subsubsection{Transport Coefficients\label{transport coefficients}}
All coefficients $\alpha,\alpha_1,\eta_g,\zeta_g$ are functions of the state variables, $u_{ij}$, $T_g$ and $\rho$. As the hydrodynamic formalism only delivers 
the structure of the dynamics, not the functional 
dependence of the  transport coefficients, these are to be obtained (same as the energy) from experiments, in a trial-and-error process. And the specification below is what we at present believe to be the appropriate ones. Generally speaking, we find 
strain dependence to be weak -- plausibly so because the strain is a small
quantity. One expand in it, keeping only the constant terms. We also expand in $T_g$, but eliminate the constant terms, because we assume granular media are fully elastic for $T_g\to0$, implying the force balance $\nabla_j\sigma_{ij}=\rho{\rm g}_i$ should reduce to the equilibrium condition, Eq~(\ref{2a-1}). Therefore we take
$\alpha,\alpha_1,\eta_g,\zeta_g,\kappa_g$ to vanish for $T_g\to0$. In addition, we
also need $\alpha,\alpha_1$ to saturate at an elevated $T_g$, such that rate-independence maybe established in the hypoplastic regime. Hence
\begin{eqnarray}\label{2c-3}
\eta_g=\eta_1T_g,\,\, \zeta_g=\zeta_1T_g,\,\,
\kappa=\kappa_1T_g,
\\\nonumber
\alpha/\bar\alpha =
\alpha_1/\bar\alpha_1={T_g}/({T_\alpha+T_g}),
\end{eqnarray} 
with $\bar\alpha,\bar\alpha_1,\eta_1,\zeta_1,\kappa_1,T_\alpha$ functions of $\rho$ only, or  the packing fraction $\phi$. Expanding $\gamma$ in $T_g$, 
\begin{equation}\label{2c-5} \gamma=\gamma_0+\gamma_1 T_g, 
\end{equation} 
we keep $\gamma_0$, because the reason that led to Eqs~(\ref{2c-3}) does not apply. More importantly, $\gamma_0$ ensures a smooth transition from the hypoplastic to the quasi-elastic regime, see Eq~(\ref{TgVs2}) below. (Although $\gamma_0=0$ in rarefied systems~\cite{luding2009}, this does not hold for denser ones.) 

The transport coefficients are also functions of $\rho$, containing especially a divergent/vanishing  part $\sim(\rho_{cp}-\rho)$. Assuming that, at $\rho=\rho_{cp}$, the plastic phenomena of stress relaxation, softening and dilatancy vanish, $T_g$ relaxes instantly, and the system is infinitely viscous, we take 
\begin{equation}\label{density-dependence}
\lambda,\,\lambda_1,\,\alpha,\,\alpha_1,\,\gamma_1^{-1},\,\eta_1^{-1}\sim\rho_{cp}-\rho.
\end{equation}  
We stand behind the temperature dependence with much more confidence than that of the density, for two reasons: First, $\rho$ is not a small quantity that one may expand in, and we lack the general arguments employed to extract the $T_g-$dependence. Second, not coincidentally, the $\rho$ dependence does not appear universal: The above dependence of $\gamma_1,\,\eta_1$ seems to fit glass beads  data, while $ \gamma_1\sim(\rho_{cp}-\rho)^{-0.5}$, $\eta_1\sim(\rho_{cp}-\rho)^{-1.5}$ appear more suitable for polystyrene beads, see~\cite{denseFlow}.

At given shear rates, $v_s=$ const, the stationary state of Eq~(\ref{2c-4}) -- characterized by $\partial_ts_g=0$,  with viscous heating  balancing $T_g$-relaxation -- is quickly
arrived at, say within $10^{-3}$ s in dense granular media, implying 
\begin{eqnarray}\label{2c-6}
{\gamma_1} \,h^2\, T_g^2=v_s^2\,{\eta_1}+v^2_{\ell\ell}\,{\zeta_1},
\\
\label{2c-5a}\text{where}\,\,\,
h^2\equiv1+\gamma_o/(\gamma_1T_g).
\end{eqnarray} 
Taking the density for simplicity as either constant or slowly changing, $v^2_{\ell\ell}\approx0$, we have a quadratic regime for small $T_g$ and low $v_s$, and a linear one at elevated $T_g,v_s$:
\begin{eqnarray}
\label{TgVs} 
T_g=|v_s|\sqrt{\eta_1/\gamma_1}\quad\,\,\text{for}\quad \gamma_1T_g\gg\gamma_0,
\\\label{TgVs2} 
T_g=v_s^2\,\,({\eta_1/\gamma_0})\quad\text{for}\quad \gamma_1T_g\ll\gamma_0.
\end{eqnarray}
As mentioned above and discussed in the next section, the linear regime is hypoplastic, in which the system displays elasto-plastic behavior and the hypoplastic model holds. In the quadratic regime, because $T_g\sim v_s^2\approx0$ is quadratically small,  the behavior is quasi-elastic, quasi-static, with slow, consecutive visit of static stress distributions. Note that we have $h=1$ in the hypoplastic regime, and $h\to\infty$ in the quasi-static one.   

Eqs~(\ref{2c-7xx},\ref{2c-8},\ref{2c-9xx}) also have a stationary solution, 
$\partial_t\Delta, \partial_tu_s=0$, in which the deformation rates $v_s,v_{\ell\ell}$ 
are compensated by the relaxation $\sim T_g$. As a result, 
$\Delta=\Delta_c,u_s=u_c$ remain constant, and with them also the 
pressure and shear stress, $P=P_c, \sigma_s=\sigma_c$. This ideally plastic behavior is the {\em critical state}. In the linear regime,
$T_g\sim|v_s|$, both $P_c$ and $\sigma_c$ are rate-independent: A higher
shear rate deforms more quickly, but the relaxation $\sim T_g$ increases
by the same amount. We shall consider the critical state in greater detail in Sec~\ref{critical state}, restating here only that since the rate-independent critical state is a motion in the linear regime, and since it is irreversible and strongly dissipative, it cannot be quasi-static.

\subsection{Summary and Three Useful Equations\label{sum}}
With the above set of equations derived, the expressions for energy density and transport coefficients in large part specified, {\sc gsh} is a fairly well-defined theory. It contains clear ramifications and provides little 
leeway for retrospective adaptation to observations. As a first step to coming to terms with its  ramifications, we examine its basic features. 

Granular rheology as observed may be divided into three shear rate regimes: {\em Bagnold} for high, {\em hypoplastic} for low, and {\em quasi-elastic} (ie.  {quasi-static}) for even lower 
ones. Fast dense flow is in the first regime, in which pressure and shear stress are proportional to shear rate squared, $p,\sigma_s\sim v_s^2$. Various elasto-plastic motions, observed especially 
in triaxial apparatuses, are in the second, rate-independent regime. The third regime is elastic.
Static stress distribution and elastic waves belong here. 
This third regime is again rate-independent, though the transition between both cannot be. 

Soil mechanics textbooks do not acknowledge the existence of a third rate regime, as they take the hypoplastic regime to be quasi-static. As mentioned above and discussed in detail in Sec~\ref{3regimes}, this cannot be right because it is irreversible and strongly dissipative. 

{\sc gsh} is constructed such that any deviation from elasticity -- encapsulated in the 
coefficients $\alpha,\alpha_1,\eta_g,\zeta_g,\kappa_g$ and the relaxation rate $\lambda
T_g,\lambda_1T_g$ of Eqs~(\ref{2c-7xx},\ref{2c-8}) -- vanishes with $T_g$.
For $T_g=0$, we have $\partial_t u_{ij}=v_{ij}\equiv\partial_t\epsilon_{ij}$, or
\begin{equation}\label{sum qe}
u_{ij}=\epsilon_{ij},\quad \sigma_{ij}=\pi_{ij},
\end{equation}
implying perfect elasticity. At very low shear rates, deviations from elasticity are quadratically small  and negligible, $T_g\sim v_s^2$. This is the {\em quasi-elastic}, or   {\em quasi-static} {regime}, because the slow motion visits a series of equilibrium, elastic states. 

When $T_g$ is more elevated, we are in the linear regime, $T_g\sim|v_s|$, see Eq~(\ref{TgVs}). Here, the full complexity of granular media emerges. Nevertheless, true to our starting assumptions on granular media: {\em two-stage irreversibility} and {\em variable transient elasticity}, three scalar equations suffice to account for most phenomena. Two account for transient elasticity, 
Eqs~(\ref{2c-7xx},\ref{2c-9xx}), and one for $T_g$, Eq~(\ref{2c-4}), 
\begin{eqnarray}
\label{2c-9}
\partial_tu_s-(1-\alpha )v_s= -\lambda T_gu_s,
\\\label{2c-7}
\partial_t\Delta +(1-\alpha)v_{\ell\ell}-\alpha_1u_sv_s
=-\lambda_1T_g\Delta, 
\\
\label{sum T_g} 
b\rho\partial_tT_g-\kappa_1T_g\nabla^2T_g=
\eta_1 v_s^2-\gamma_1 h^2(T_g^2-T_a^2). 
\end{eqnarray}
Some simplifications and one modification were made to arrive at these equations: (1)~The gradient of $T_g$ is assumed to be small, and linearized in; all other variables were taken to be uniform. (2)~$T_g$'s convective term is taken to be negligible. 
(3)~An extra source term $\gamma_1 h^2T_a^2$ is added to account for an ambient temperature $T_a$ -- external perturbations such as given by a sound field or by tapping. 
Generally speaking, any source mechanism contributing to $T_g$ is already included in the expression without $T_a$. For instance, given a sound field -- generated either by loudspeakers or tapping -- there is the term on the right hand side 
of Eq~(\ref{2c-4} ),  $\zeta_1 (v_{\ell\ell}^{sound})^2$, where $v_{\ell\ell}^{sound}$ is the fast varying compressional rate of the sound field. Coarse-graining it, we may set 
\begin{equation}\label{v_T}
\langle\zeta_1 (v_{\ell\ell}^{sound})^2\rangle \equiv\gamma_1h^2T_a^2\equiv \eta_1 v_T^2,
\end{equation}
to quantify this contribution, either in terms of the ambient temperature $T_a$, or a shear rate $v_T$ needed to produce this $T_a$. Adding this term is a convenient short cut to account for a general perturbation, for an ambient temperature without specifying the cause.

The information on the elastic strain $\Delta(t), u_s(t)$ is, for given density, identical as that of the elastic stress $P_\Delta(t), \pi_s(t)$, because they are always given by the hyper-elastic relation, Eq~(\ref{1-1}), or by Eqs~(\ref{2b-2a}, \ref{2b-2b}) for the elastic energy Eq~(\ref{2b-2}). The total stress includes the seismic pressure $P_T$ and the viscous contributions (of which the compressional one is neglected below). We write (assuming that the lowest order terms $\sim \Delta^{1.5}$ dominate and neglecting the term  $-\bar\alpha_1u_sP_\Delta$, of order $\Delta^{2.5}$) 
\begin{eqnarray}\label{sum stress df}
P&=&(1-\bar\alpha)P_\Delta+{\textstyle\frac12} T_g^2\, a\, \rho^2 \,b/(\rho_{cp}-\rho),
\\\nonumber
\sigma_{s}&=&(1-\bar\alpha)\pi_s-\eta_1T_gv_s.
\end{eqnarray}
[Given Eq~(\ref{2b-2}), $\pi^*_{ij}$, $u^*_{ij}$ and $\sigma^*_{ij}$ are
colinear.]  In the linear regime, $T_g\sim|v_s|$, the elastic terms $P_\Delta,\pi_s$ are (as discussed at the end of the last section) rate-independent, while $P_T\sim T_g^2$ and $\eta_gv_s=\eta_1T_gv_s$ are quadratic in $v_s$. 
So both may be written as $e_1+e_2v_s^2$, implying a quadratic
dependence on the rate for $e_2v_s^2\gg e_1$, and rate-independence for  $e_2v_s^2\ll e_1$. The first 
limit may be identified with the {Bagnold regime}, the second with the {hypoplastic} one. For the ease of future references, we note for the hypoplastic regime:
\begin{equation}\label{sum stress hp}
P=(1-\bar\alpha)P_\Delta, \quad \sigma_{s}=(1-\bar\alpha)\pi_s.
\end{equation}• 
This ends the brief presentation of {\sc gsh}.


\section{The Quasi-Elastic Regime\label{quasi elastic motion}}

\subsection{Quasi-Elastic versus Hypoplastic Regime\label{3regimes}} 

Textbooks on soil mechanics take granular motion in the hypoplastic  regime -- say the approach to the critical state -- to be quasi-static. We do not believe this is right,  because although slow and rate-independent, it is also strongly dissipative and irreversible. 

Quasi-static motion is never dissipative. Think of sound propagation in any system (such as Newtonian liquid, elastic media or liquid crystals), where the velocity is an order in the frequency lower than the damping. So sound waves are less damped the smaller the frequency is. This is a general feature: Changing a state variable: $A$ slowly, dissipation vanishes with the rate of change $\partial_tA$. The motion is therefore rate-independent in the very slow limit, in which dissipation may be neglected. We call it  {\em quasi-static} because the system is at this rate visiting static, equilibrium states consecutively.  

Granular systems are both dissipative and rate-independent in the hypoplastic regime. Because of rate-independence, reactive and dissipative terms are of the same order in  the frequency, and comparable in size. (They are exactly equal in the critical state.) If there were only the hypoplastic regime, elastic waves would always be overdamped. Since this is not the case, there must be a dissipation-free, quasi-static one that we term quasi-elastic.  

A frequent suggestion is to take a small incremental strain (such as given in an elastic wave) to be elastic and free of dissipation, but a large one as elasto-plastic and dissipative. We believe this is the wrong way out, because it is (taken literally) illogical and incompatible with the notion of a quasi-static motion: Starting from a static state of given stress, and applying a small incremental strain that is elastic, the system is again in a static state and an equally valid starting point. The next small increment must therefore also be purely elastic. Many consecutive small increments yield a large change in strain, and if the small ones are not dissipative, neither can their sum be. 

In {\sc gsh}, it is the strain rate rather than strain amplitude that decides whether the system is elastic or elasto-plastic.  Of course,  small strain increments achieved with
a higher but short lasting shear rate will indeed provoke elastic responses, if
$T_g$ does not have time to get to a sufficiently high value to induce
plastic responses. Furthermore, the mere existence of a quasi-static, quasi-elastic regime does not imply that it is also easily observable.

To be specific, we quote a few numbers, though these are at best educated guesses. Aside from the lack of
unambiguous experimental data, circumstances are
usually complicated by density or pressure dependence~\cite{midi}.  We believe, the Bagnold regime starts at shear rates of one or two hundred Hz, the hypoplastic regime is say between $10^{-3}-1$Hz, and quasi-elastic regime lies possibly below $10^{-5}$Hz.

Finally, we note that backtracing of the stress curve $\hat\sigma(t)$ when reversing the strain, $\hat\epsilon(t)\to\hat\epsilon(-t)$, occurs only in
the quasi-elastic regime, not the hypoplastic one. (We use a hat
to indicate a tensor.)  The stress is a function of the
elastic strain, $\hat\sigma=\hat\sigma(\hat u)$. Reversing $\hat u(t)$ will
always backtrace $\hat\sigma(t)$. But only in the quasi-elastic regime may
we identify $\hat u(t)=\hat\epsilon(t)$. Failure to backtrace at hypoplastic rates are not evidence of ``history dependence."

\subsection{A Steep Stress-Strain Trajectory\label{A Steep Stress Trajectory}}
As discussed above, in the quadratic regime of very slow shear rates,  $T_g\sim |v_s|^2\to0$, the granular temperature is so small that the system is essential elastic, moving from one elastic, equilibrium state to a slightly different one. This is the reason we call it {\em quasi-elastic}, or {\em quasi-static}. Because $\hat\sigma\to\hat\pi$ and $\partial_t\hat u\to\partial_t\hat\epsilon=\hat v$, the change of the the shear stress $\sigma_s$ is well approximated by the (hyper-) elastic relation, 
\begin{equation}\label{3a-1}
\partial_t\sigma_{ij}=\frac{\partial\sigma_{ij}}{\partial
u_{k\ell}}\partial_t u_{k\ell} =\frac{\partial\pi_{ij}}{\partial
u_{k\ell}}\partial_t\epsilon_{k\ell}=-\frac{\partial^2 w}{\partial
u_{ij}\partial u_{k\ell}}v_{k\ell}. 
\end{equation} 
Shearing a granular medium at quasi-elastic rates, the result will be a trajectory $\hat\sigma(\hat\epsilon)$ that is much steeper than in experiments at hypoplastic rates, such as observed during an approaching to the critical state. The gradient is given directly by the stiffness constant ${\partial^2 w}/{\partial \hat u^2}$, and possibly three to four times as large as the average between loading and unloading at hypoplastic rates [because Eq~(\ref{2c-8}) lacks the factor of $(1-\alpha)$]. This goes on until the system reaches a yield surface of the elastic energy, one of those discussed in Sec~\ref{yield surfaces}. We expect the system to form shear bands at this point, see Sec~\ref{sb}. The critical state will not be reached. Reversing the shear rate in between will retrace the function $\hat\sigma(t)$. 

\subsection{Soft Springs versus Step Motors\label{soft springs}}
\begin{figure}[t] \begin{center}
\includegraphics[scale=0.6]{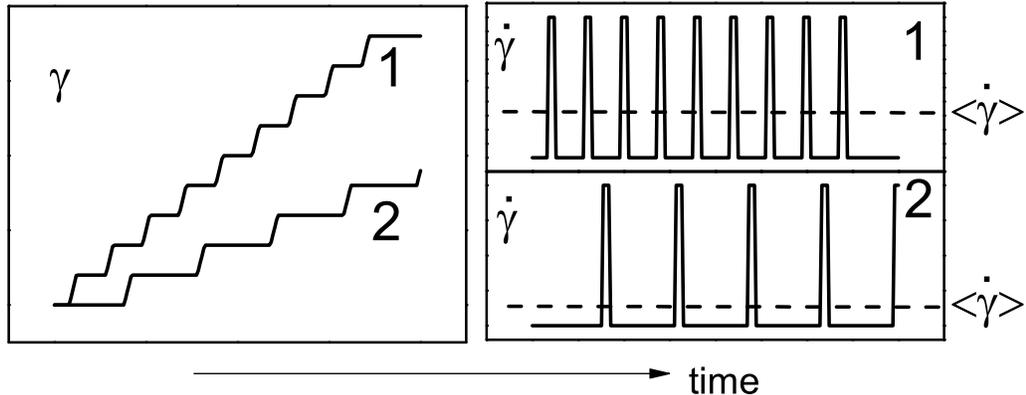}
\end{center}
\caption{Why it is hard to observe the quasi-elastic regime if step motors are used, see text.\label{StepMotor}} 
\end{figure} 

Quasi-elastic behavior has not been observed in triaxial apparatus, even at the lowest rates. This maybe because they are simply not slow enough. Quite probably though, this is also due to the wide usage of step motors in these appliances. Plotting its shear 
rate versus time, $\dot\gamma(t)$, different shear rates are approximately given as depicted by the two curves of Fig~\ref{StepMotor}. Although the curves have different average rates $\langle\dot\gamma\rangle$, the time-resolved, maximal rates  $\dot\gamma_M$ are identical. And if the time span of  $\dot\gamma_M$ is long enough for $T_g$ to respond, and $\dot\gamma_M$ is high enough for the system to be in the linear regime, $T_g\sim \dot\gamma_M$, the system will display consecutive hypoplastic behavior in both cases, irrespective of  the average rate $\langle\dot\gamma\rangle$. 

We suggest two ways here to circumvent this difficulty, both by fixing the stress rate at low $T_g$: As discussed in the last section, a given stress rate has a high shear rate at elevated $T_g$ and a low one at vanishing $T_g$.  First is slowly tilting an inclined plane supporting a layer of grains. In such a situation, the shear rate remains very small, and the system starts flowing only when a yield surface is breached. In contrast, employing a feedback loop in a triaxial apparatus to maintain a stress rate would not work well, because the correcting motion typically has strain rates that are too high.

A second method is to insert a very {soft spring}, even a rubber band, between the granular
medium and the device moving at a given velocity $v$ to deform it. If the spring is softer by a large factor $a$ than the granular medium (which is itself rather soft), it will absorb most of the displacement, leaving the granular medium deforming at a rate smaller by
the same factor $a$ than without the spring. In other words, the soft spring serves as a ``stress reservoir'' for the granular medium. The same physics applies when the feedback loop is connected via a soft spring. Little $T_g$ is then excited, see for instance the experiment discussed in Sec~\ref{aging}.

\section{The Hypoplastic Regime\label{hypoplastic motion}}

Hypoplastic motion occurs at an elevated $T_g\sim|v_s|$, in what we have named the linear regime. It is  {\em rate-independent} for given, constant strain rates, in the sense that the increase in the stress $\Delta\sigma_{ij}$ depends only on the increase in the strain,
$\Delta\epsilon_{ij}=\int v_{ij}{\rm d}t$, not how fast it takes place. We call this regime hypoplastic because this is where the {\em hypoplastic model} holds, a state-of-the-art engineering theory~\cite{kolymbas1} that we shall consider in Sec~\ref{Hypoplasticity}.

\subsection{Load and Unload\label{Load and Unload}} 

In the hypoplastic regime, for given shear rate $v_s$, the granular temperature relaxes quickly to its stationary value  $T_g=|v_s|\,\sqrt{\eta_1/\gamma_1}$. Inserting this into
Eqs~(\ref{2c-9}, \ref{2c-7}), we arrive at
\begin{eqnarray}\label{3b-2}
\partial_t\Delta=v_s\,\alpha_1u_s-|v_s|\,\Lambda_1\Delta, 
\\\label{3b-3}
\partial_tu_s=v_s\,(1-\alpha )-|v_s|\,\Lambda u_s, 
\\\label{3b-4}
\Lambda\equiv\frac{\lambda}{h}\sqrt{\frac{\eta_1}{\gamma_1}}\equiv\Lambda_1\frac{\lambda}{\lambda_1}\sim
(\rho_{cp}-\rho), \end{eqnarray} 
which are explicitly rate-independent for $\alpha=\bar\alpha,\alpha_1=\bar\alpha_1$. 
The last equation is a result of inserting the density dependence of Eqs~(\ref{density-dependence}) and indicates that relaxation of the elastic strain becomes slower at higher density, and stops at the close-packed density $\rho_{cp}$, where the system is essentially elastic. We take  $\Lambda\approx3.3\Lambda_1$, as compressional relaxation is typically slower than shear relaxation~\cite{granL3}. 

\begin{figure}[tbh] \begin{center}
\includegraphics[scale=0.6]{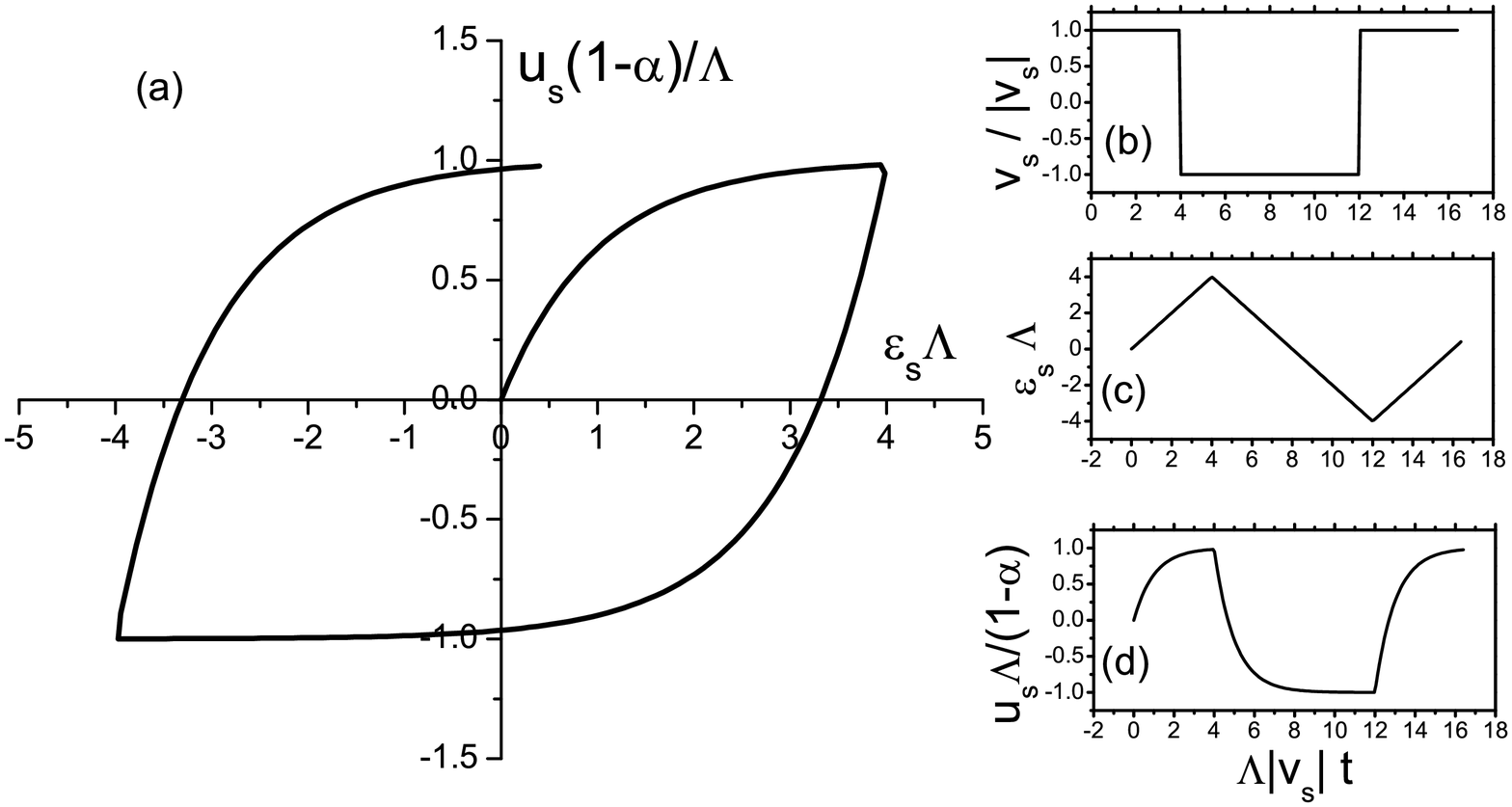}
\end{center}
\caption{\label{fig2} The hysteretic change of the shear stress ($\sim u_{s}$) with the strain, as accounted for by Eq~(\ref{3b-3}). The sign of the shear rate $v_{s}(t)$ is given in (b), the shear deformation $\varepsilon _{s}=\int_{0}^{t}v_{s}(t^{\prime })dt^{\prime }$ in
(c). Inset (d) is the the temporal evolution of $u_{s}$. } 
\end{figure} 

In this form, it is obvious that loading ($v_s=|v_s|>0$) and unloading ($v_s=- |v_s|<0$) have different slopes: $\partial_tu_s/v_s=(1-\alpha)\mp(\Lambda u_s/h)$.
This phenomenon is referred to as {\em
incremental nonlinearity} in soil mechanics, and the reason why no backtracing takes place under reversal of shear rate: Starting from isotropic stress, $u_s=0$, see Fig~\ref{fig2}, the gradient is at first $(1-\alpha)$, becoming smaller as $u_s$ grows, until it
is zero, in the stationary case $\partial_tu_s/v_s=0$. Unloading now, the
slope is $(1-\alpha)+(\Lambda u_s/h)$, steeper than it has ever been. It is again 
$(1-\alpha)$ for $u_s=0$, and vanishes for $u_s$ sufficiently negative, see Fig~\ref{fig2}. Same scenario holds for $\partial_t\Delta/v_s$. 

Clearly, only the stress $P,\sigma_s$ are measurable, not
$\Delta, u_s$. The former is calculated employing Eq~(\ref{sum stress hp}) when
the latter is given. The resultant expressions can be complicated
(especially if the pressure is held constant instead of the density, see
Sec~\ref{critical state}), but the basic physics remains the same -- an
illustration of why $u_{ij}$ is the better state variable.

In systematic studies employing discrete numerical simulation, Roux and coworkers have 
obtained great knowledge about the mesoscopic physics on intergranular scales, see 
eg.~\cite{roux}. And they were especially able to distinguish between two types of strain, I 
and II, complete with two regimes in which either dominates. However, although  
type I strain may clearly be identified as our state variable $u_{ij}$, one needs to be aware that regime~I  is not necessarily {\em quasi-static}, or {quasi-elastic} as considered in Sec~\ref{quasi elastic motion}. The difference is: The relaxation term may be temporarily small at hypoplastic shear rates, say because $u_s$ or $\rho_{cp}-\rho$ are, see Eqs~(\ref{3b-2}, \ref{3b-4}), they do not stay small if one wanders in the variable space. At quasi-elastic rates, deformation are always free of dissipation.

\subsection{The Critical State\label{critical state}} 
\subsubsection{Stationary Elastic Solution\label{Stationary Elastic Solution}}
\begin{figure}[t] \begin{center}
\includegraphics[scale=0.27]{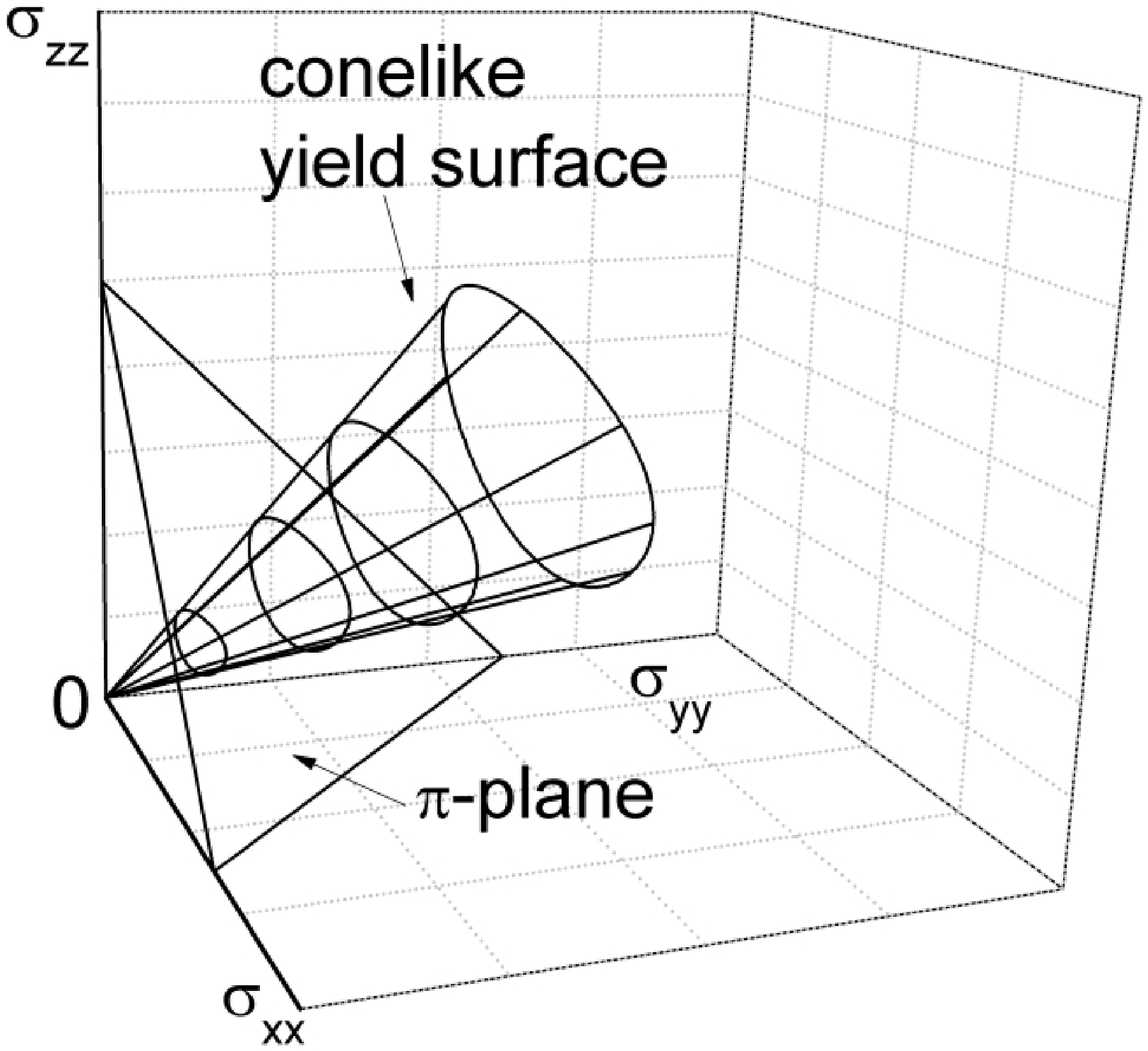}
\includegraphics[scale=0.26]{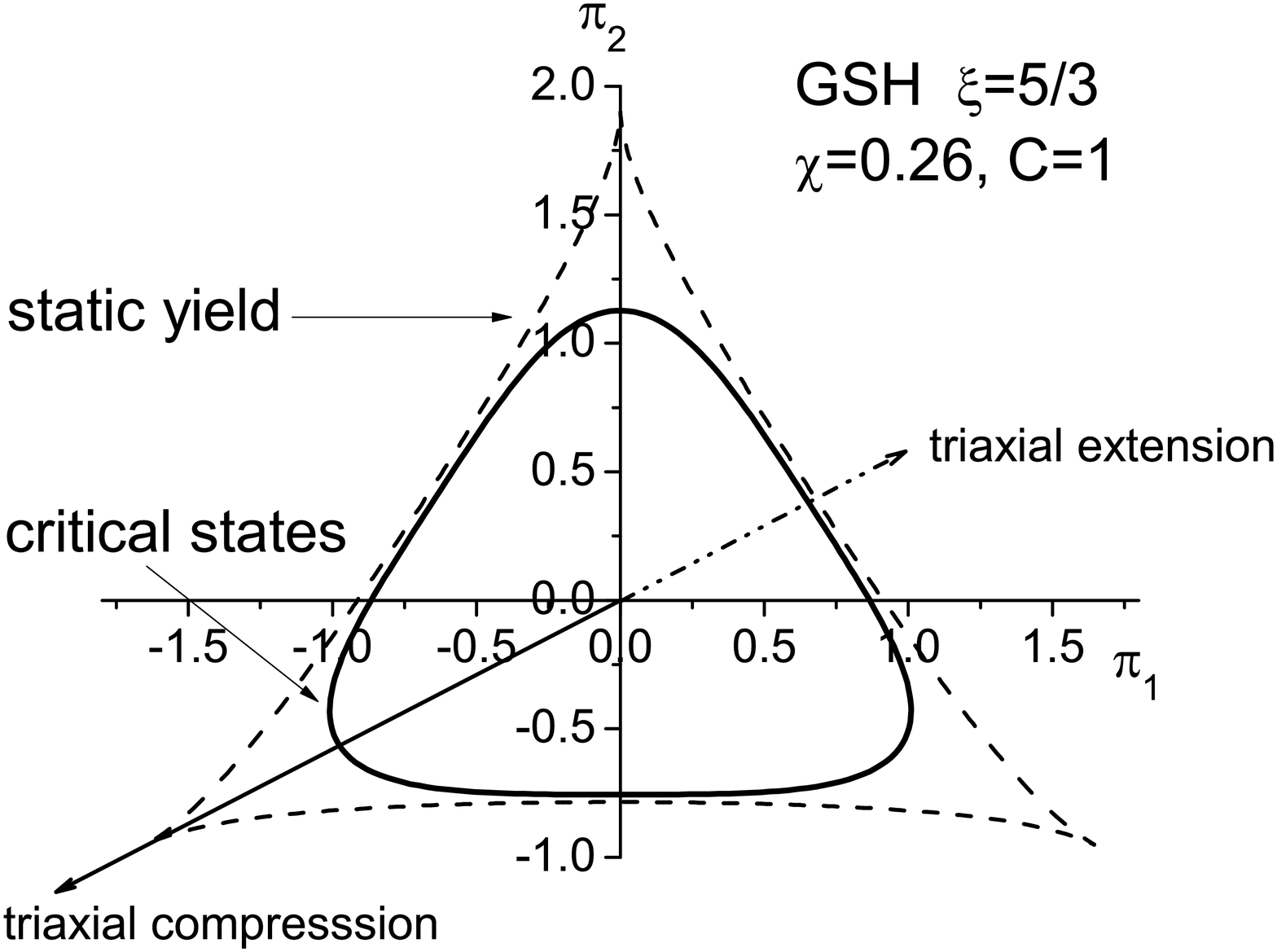}
\end{center}
\caption{\label{yield-cs-pai} Loci of static yield surface and the critical states, calculated employing the more general energy of Eq~(\ref{2b-6}). Left: in the space spanned by the three stress eigenvalues, $\sigma_1, \sigma_2, \sigma_3$; right: in the $\pi$-plane of constant pressure, $P\equiv\sigma_1+\sigma_2+\sigma_3$, where $\sqrt{2}\pi _{1} \equiv(\sigma _{3}-\sigma _{2})/P$, 
$\sqrt{6}\pi _{2} \equiv(2\sigma _{1}-\sigma _{2}-\sigma _{3})/P$. }
\end{figure}

When there is complete compensation of the shear rate $\sim v_s$ and the relaxation
$\sim T_g$, the stationary solution of Eqs~(\ref{3b-2},\ref{3b-3}) for the elastic strain 
$u_{ij}$ holds. It is generally called the {\em critical state}, see~\cite{critState}, and may be considered ideally plastic, because a shear rate does not lead to a stress 
increase. Setting $\partial_t\Delta, \partial_tu_s=0$ in Eqs~(\ref{2c-7},\ref{2c-9}), we obtain the somewhat more general expressions (useful for Sec~\ref{external perturbation}), 
\begin{equation}\label{3b-3a} u_c=\frac{1-\alpha}{\lambda}\frac{v_s}{T_g}=\pm\frac{1-\alpha}{\Lambda},
\quad \frac{\Delta_c}{|u_c|}=\frac{\alpha_1}{\lambda_1}\frac{|v_s|}{T_g}=\frac{\alpha_1}{\Lambda_1}.
\end{equation} 
From Eq~(\ref{2c-8}), the colinearity of the critical strain and rate, $u^*_{ij}|_c/|u_c|=v^*_{ij}/|v_s|$, is easy to see. 
In the hypoplastic regime (for $h=1, \alpha=\bar\alpha,\alpha_1=\bar\alpha_1$),  $u_c,\Delta_c$ depend only on the density and is rate-independent. 
The critical stress is given by inserting $u_c,\Delta_c$ into Eqs~(\ref{sum stress hp}), 
\begin{eqnarray}\label{3b-3b} P^c_\Delta&\equiv&P_\Delta(\Delta_c,
u_c)=\sqrt\Delta_c({\cal B}\Delta_c+{\cal A} {u_c^2}/{2\Delta_c}),
\\\label{3b-3c}
\pi_c&\equiv&\pi_s(\Delta_c, u_c)=-2{\cal A}\sqrt\Delta_c\, u_c, 
\\\label{3b-3cA}
{P^c_\Delta}/{\pi_c}&=&({{\cal B}}/{2{\cal
A}}){\Delta_c}/{u_c}+{u_c}/{4\Delta_c}, 
\\\label{3b-4a}
P_c&=&(1-\bar\alpha)P^c_\Delta,\quad
\sigma_c=(1-\bar\alpha)\pi_c.
\end{eqnarray} 
The critical ratio $\sigma_c/P_c$ -- same as the Coulomb yield of Eq~(\ref{2b-3}) -- is also frequently associated with a friction angle. Since one is relevant for vanishing $T_g\sim v_s^2\to0$, while the other requires an elevated $T_g\sim |v_s|$, it is appropriate to
identify one as the static friction angle, and the other as the
dynamic one. The dynamic friction angle is always smaller than
the static one,  see Fig~\ref{yield-cs-pai}, because the critical state is elastic,
and must stay below Coulomb yield, 
\begin{equation}\label{3b-5} \Lambda_1/\bar\alpha_1<\sqrt{2{\cal B/A}}.
\end{equation}

Textbooks on soil mechanics frequently mention that the friction angle is
essentially independent of the density -- although they do not, as a rule,
distinguish between the dynamic and the static one, cf. Sec~\ref{yield
surfaces}. We assume, for lack of more discriminating information, that both are. Therefore, we take $\alpha_1\sim(\rho_{cp}-\rho)$, because $\Lambda_1$ also does, see Eq~(\ref{3b-4}).
Quite generally, we note that accepting the density dependence of Eqs~(\ref{density-dependence}), we have $\Delta_c, u_c$ being monotonically increasing functions of $1/(\rho_{cp}-\rho)$. Same holds for $P_c,\sigma_c\sim{\cal B}$, though ${\cal B}$'s density dependence make the increase slightly faster.

\subsubsection{Approach to the Critical State\label{approach critical}} 
Solving Eqs~(\ref{3b-2},\ref{3b-3}) for $u_s(t),\Delta(t)$, at constant
$\rho, v_s$ and $h=1, \alpha=\bar\alpha,\alpha_1=\bar\alpha_1$, with the initial conditions: $\Delta=\Delta_0, u_s=0$, the approach to the critical state is given as
\begin{eqnarray}\label{3b-6}
u_s&(t)&=u_c(1-e^{-\Lambda\varepsilon_s}),\quad \varepsilon_s\equiv v_st,
\\\nonumber \Delta&(t)&=\Delta_c(1+f_1\,e^{-\Lambda
\varepsilon_s}+f_2e^{-\Lambda_1\varepsilon_s}), \\\nonumber
f_1&=&\frac{\Lambda_1}{\Lambda-\Lambda_1},\quad
f_2=\frac{\Delta_0}{\Delta_c}-\frac{\Lambda}{\Lambda-\Lambda_1}.\end{eqnarray}
showing that the approach is a simple exponential decay for $u_s$, and a sum of two decays for $\Delta$. It is useful, and demystifying, that a simple, analytical solution in terms of the elastic strain exists. Because $\Lambda\approx3.3\Lambda_1$ [see the
remarks below Eq~(\ref{3b-4})], the decay of $u_s$ and $f_1$ are 
faster than that of $f_2$. Note $f_2$ may be
negative, and $\Delta(t)$ is then not monotonic. The associated pressure
and shear stress are those of Eqs~(\ref{3b-3b},\ref{3b-3c},\ref{3b-3cA},\ref{3b-4a}). For negative $f_2$, neither the pressure nor the shear stress is monotonic. 

For the system to complete the approach to the critical state, the yield surface must not be breached during the non-monotonic course of the shear stress. If it happens, shear bands will form, see Sec~\ref{sb} below, and the uniform critical state will not be reached.

\subsubsection{Pressure-Controlled Approach\label{pressure approach}}
\begin{figure}[t] 
\begin{center}
\includegraphics[scale=0.6]{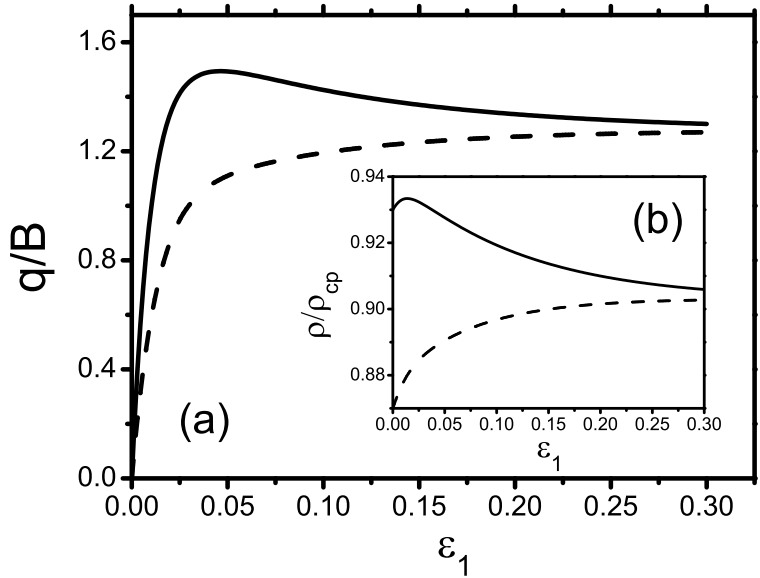}
\end{center}\caption{\label{fig3} Triaxial test curves computed with GSH for a loose (dashed) and dense initial density -- the axial stress $\sigma_2=\sigma_3$ is being hold constant, with $q\equiv\sigma_1-\sigma_2$.}
\end{figure}

Frequently, the critical state is not approached at constant density (ie.
volume), but at constant pressure $P$ (or a stress eigenvalue $\sigma_i$). The circumstances are then more complicated. As $\Delta,u_s$ approach
$\Delta_c,u_c$, the density compensates to maintain
$P(\rho,\Delta,u_s)$. Along with $\rho$, the coefficients
$\alpha,\alpha_1,\Lambda,\Lambda_1$, all functions of $\rho$, also change
with time. In addition, with $\rho$ changing, compressional flow
$v_{\ell\ell}=-\partial_t\rho/\rho$ no longer vanishes (though it is still small). Analytic solutions
do not seem feasible now, but numerical ones are.  Reassuringly, our result is a perfect textbook illustration, see Fig~\ref{fig3}.

Generally speaking, we have four scalar state variables: $\rho,T_g,u_s,\Delta$, each with an equation of motion that depends on the rates $v_s,v_{\ell\ell}$ and  the variables themselves. In addition, 
$P,\sigma_s$ are functions of $\rho,u_s,\Delta$. In the last section, both rates were given, 
$v_{\ell\ell}=0$, $v_s=$ const. As a result, we have $\rho=$ const, while $\Delta(t)$ and $u_s(t)$ were calculated taking the coefficients $\alpha(\rho),\alpha_1(\rho),\Lambda(\rho),
\Lambda_1(\rho)$ as constant. The stress components were then obtained as dependent 
functions. A pressure-controlled experiment means that only the shear rate $v_s$ is given. Holding $P(\rho,u_s,\Delta)=$ const (or analogously  $\sigma_1$) implies
the density $\rho$ (and with it also $v_{\ell\ell}=-\partial_t\rho/\rho$) is a dependent function, $\rho=\rho(P,u_s,\Delta)$. Now, in the equations of motion for $u_s$ and $\Delta$, one first eliminates
$v_{\ell\ell}$ employing  $v_{\ell\ell}=-\partial_t\rho/\rho$, then eliminates both $\partial_t\rho/\rho$ and the $\rho$-dependence
of $\alpha(\rho),\alpha_1(\rho),\Lambda(\rho),\Lambda_1(\rho)$ employing
$\rho=\rho(P,u_s,\Delta)$. This changes the differential equations -- which are then solved numerically.

Many well-known features of Fig~\ref{fig3} can be understood assuming the solutions of Eq~(\ref{3b-6}) remain valid, say because the initial density is close to 
the critical one, hence it does not change much during the approach to the critical state. As a 
result, we may approximate $\alpha(\rho),\alpha_1(\rho),\Lambda(\rho),\Lambda_1(\rho)$ as constant, and take $v_{\ell\ell}\approx0$.  In addition, we shall assume $\Lambda\gg\Lambda_1$, or ${\Lambda}/({\Lambda-\Lambda_1})\approx1$, instead of  $\,\approx1.5$. Then $f_2$ has
the same sign as $\Delta_0-\Delta_c$. Now consider the initial stress, 
$\sigma_s=0$ and $P\sim{\cal B}(\rho_0)\Delta_0^{1.5}$. For $P$ given and ${\cal B}(\rho)$ a monotonically increasing function of $\rho$, the pair
$\Delta_0-\Delta_c$ and $\rho_0-\rho_c$ have reversed signs. Therefore, we have a monotonic change of density for $f_2>0$, $\Delta_0>\Delta_c$, $\rho_0<\rho_c$,  and non-monotonic change otherwise. At
the beginning, the faster relaxation of $f_1$ dominates, so $\Delta$ always
decreases, and $\rho$ always increases, irrespective of $\rho_0$. After
$f_1$ has run its course, $\rho$ goes on increasing for the low density
case but switches to decreasing for the high-density case, until the
critical state is reached. These are the cases respectively referred to as
{\em contractancy} and {\em dilatancy}. This seems to be indeed what happens, although we do not have  $\Lambda\gg\Lambda_1$ in reality.

The shear stress $\sigma_s\sim \sigma_1-\sigma_2$ always increases first
with $u_s$, until $u_s$ is close to $u_c$. The subsequent behavior depends on
what $\Delta$ does. With $P\sim{\cal B}(\rho_0)\Delta_0^{1.5}$ given, $\sigma_s\sim{\cal B}\Delta^{0.5}\sim P/\Delta$ keeps growing if $\Delta$ decreases [loose
case, $f_2>0$], but becomes smaller again, displaying a peak, if $\Delta$ grows [dense
case, $f_2<0$]. 

As mentioned, during the approach to the critical case, the yield surface
may be breached. The system will then interrupt its approach to the
critical state, and develop shear band instead, see Sec~\ref{sb}.

\subsubsection{External Perturbation or Ambient Temperature  \label{external perturbation}}
If one perturbs the system, say by exposing it to a standing sound field, or more simply, by tapping it 
periodically, such as in a recent experiment~\cite{vHecke2011}, the critical state is strongly modified. This is the kind of  games/explorations physicists play. To engineers, it may seem less than serious, even a bit frivolous, but 
it does time and again lead to useful insights. In~\cite{vHecke2011}, a surprising rate-dependence of the critical shear stress was observed. The stress decreases with the tapping 
amplitude, and increases with the shear rate, such that the decrease is compensated at 
higher rates. Clearly, engineering theories, be they elasto-plastic or hypoplastic, that build in rate-independence from the start, could not possibly 
account for this observation. {\sc gsh}, on the other  hand,  should be able to, if it indeed 
provides a wide-range qualitative account of granular behavior. 

The consideration of the critical state in the previous three sections takes any granular temperature $T_g$ to be a 
result of the given shear rate, hence  $T_g=|v_s|\,
\sqrt{\eta_1/\gamma_1}$, or  Eq~(\ref{TgVs}), holds. This is no longer the case here, as sound field or tapping will in addition contribute to $T_g$. We have called this portion of $T_g$ the ambient temperature $T_{a}$, or the shear rate $v_T\equiv T_a\sqrt{\gamma_1/\eta_1}$ needed to produce this $T_a$. In Eq~(\ref{sum T_g}), taking $\partial_tT_g,\nabla_iT_g=0$ for a stationary and uniform system, we have (for $h=1$)
\begin{equation}\label{3b-7}
\frac{T_g}{v_s}=\sqrt{\frac{\eta_1}{\gamma_1}}\,\,\sqrt{1+\frac{v_T^2}{v_s^2}}.
\end{equation}
Inserting this into Eq~(\ref{3b-3a}), we find the perturbed critical strain and stress: $\bar u_c, \bar\Delta_c, \bar\sigma_c\sim \bar u_s\sqrt{\bar \Delta}$, given as
\begin{equation}\label{3b-8}\frac{\bar u_c^2}{u_c^2}=
{\frac{\bar \Delta_c}{\Delta_c}}={\frac{\bar \sigma_c}{\sigma_c}}=\frac1{1+{ v_T^2}/{v_s^2}}.
\end{equation}
If there is no tapping, $T_a\sim v_T=0$, we retrieve the unperturbed values, $\bar u_c=u_c, \bar\Delta_c=\Delta_c, \bar\sigma_c=\sigma_c$. With tapping,  $\bar u_c, \bar\Delta_c, \bar\sigma_c$ decrease for increasing $T_a\sim v_T$, and increase with increasing shear  rate $v_s$, as observed.

\subsection{Stress-Controlled Experiments\label{aging}} 
\subsubsection{Long-Lived Temperature and Diverging Strain}
In Sec~\ref{critical state}, only rate-controlled experiments, with $v_s$ given, were considered. Employing Eqs~(\ref{2c-7},\ref{2c-9},\ref{sum T_g}), we found that
\begin{itemize}
\item the granular temperature quickly becomes a dependent quantity,  $T_g=|v_s|\,
\sqrt{\eta_1/\gamma_1}$, with {\sc gsh} reducing to the hypoplastic model.
\item  The exponential relaxation of $\Delta,u_s$ reproduces the approach to the critical state. 
\end{itemize}
In this section,  we examine what happens if we instead hold
the shear stress $\sigma_s$  constant. If $T_g=0$, the system stays static, $\sigma_s=$ const, and there 
is no dynamics at all. If $T_g$ is initially elevated, $u_s$ relaxes, and with it also the stress $\sigma_s$. 
Maintaining a constant  $\sigma_s$ (or similarly, a constant $u_s$) therefore requires a compensating shear 
rate $v_s$. 
As long as $T_g$ is finite, $v_s(t)$ will accumulate, resulting in a growing shear strain 
$\epsilon_s(t)=\int v_s{\rm d}t$. As we shall see, for $u_s$ close to its critical value $u_c$,  the characteristic time of $T_g$ is $\sim(1-u^2_s/u^2_c)^{-1}$ and long. Adding in the fact that $T_g$ relaxes algebraically slow rather than exponentially fast, the accumulated shear strain can be expected to be very large. 

In a recent experiment, Nguyen et al.~\cite{aging} pushed the system 
to a certain shear stress at a given and fairly fast rate, producing an elevated $T_g$. Then, switching to maintaining the shear stress, they observed a large total strain  
$\epsilon_s(t)$ that appears to diverge logarithmically. The authors referred to this phenomenon 
as creeping, and took it to be a {compelling evidence} that the slow motion of the experiment contains a 
dynamics and cannot be quasi-static. we note that this conclusion sits well with a basic contention of {\sc 
gsh}, that what is usually taken as quasi-static motion is in fact hypoplastic, with an elevated $T_g$, see 
Sec~\ref{3regimes}.  

This experiment may in principle be accounted for by the equations of {\sc gsh} in the hypoplastic 
regime, though this -- due to its  complicated geometry 
and highly nonuniform stress distribution -- requires solving a set of nonlinear partial differential equations. Here, we only engage in a qualitative consideration of shear-stress controlled experiments in the hypoplastic regime, assuming uniform variables. Also, we assume for algebraic simplicity that it is the elastic shear strain $u_s$ that is being kept constant, not the shear stress $\sigma_s\sim\sqrt\Delta\,u_s$. 
The relevant equations are still Eqs~(\ref{2c-7},\ref{2c-9},\ref{sum T_g}). 

At the beginning, as the strain is being ramped 
up to $u_s$ employing a constant rate $v_1$, the granular temperature acquires the elevated value $T_0=(v_1/h)\sqrt{\eta_1/\gamma_1}$. Starting at $t=0$, $u_s$ is being held constant. Setting $\partial_tu_s=0$ in  Eq~(\ref{2c-9}),  the shear rate needed to compensate the stress relaxation is (see also Eq~(\ref{3b-3a}))
\begin{equation}\label{3b-7a}  
v_s=\frac{\lambda T_gu_s}{1-\alpha}=T_gh\frac{u_s}{u_c}\sqrt{\frac{\gamma_1}{\eta_1}}.
\end{equation}
(Note that with $v_T\equiv T_g\sqrt{\gamma_1/\eta_1}$ as the shear rate needed to produce $T_g$ at given rates, we have the relation $v_s/v_T=u_s/u_c$.) Inserting Eq~(\ref{3b-7a}) into Eqs~(\ref{2c-7},\ref{sum T_g}) yields
\begin{eqnarray}\label{3b-8a} 
\partial_t\Delta&=&-\lambda_1T_g(\Delta-\Delta_\infty),\quad \Delta_\infty\equiv{\Delta_c} u_s^2/u_c^2
\\\label{3b-9} 
\partial_t T_g&=&-(\gamma_1/b\rho)[1-u_s^2/u_c^2]\,h^2T_g^2.
\end{eqnarray}
These equations may be solved analytically, if the coefficients are constant, which they are if the density is. The pressure $P(t)\sim\Delta(t)$ will change with time. This is what we consider here.  Next, in Sec~\ref{sapc}, we take the pressure as a constant, implying time-dependence of density and coefficients. Then, as with the critical state considered in Sec~\ref{pressure approach}, a general solution is possible only by numerical methods.   

The first equation accounts for the relaxation of $\Delta$, from both below and above $\Delta_\infty$. The relaxation is faster the more elevated $T_g$ is. 
Writing the second equation as $\partial_t T_g=-AT_g^2$, setting  $h=1$, and employing the initial condition $T_g=T_0$ at $t=0$, we obtain the solution
\begin{equation}\label{3b-10} 
T_g=\frac{T_0}{1+{A}{T_0}t}, \quad A\equiv\frac{\gamma_1}{b\rho}
\left(1-\frac{u_s^2}{u_c^2}\right).
\end{equation}
Because of Eq~(\ref{3b-7a}), we may rewrite the solution as 
\begin{equation}\label{3b-11} 
v_s=\frac{v_0}{1+ Cv_0 t}, \quad \frac CA\equiv \frac{1-\alpha}{\lambda u_s},
\end{equation}
which implies a slowly growing total strain
\begin{equation}
\epsilon_s-\epsilon_0\equiv\int_0^tv_s{\rm d} t=\ln(1+Cv_0t)/C.
\end{equation}
As $v_s$ diminish, it will eventually enter the quasi-elastic regime, 
$\gamma_1h^2T_g^2\to \gamma_0T_g$, and the last bit of $v_s$ relaxes exponentially. Therefore, $\epsilon_s$ does not really diverge.

Assuming a large rate to ramp up the stress, the initial value for the granular temperature $T_0$ is also large. This will quickly let $\Delta$ be fully relaxed, $\Delta=\Delta_\infty$. Fixing 
$u_s=u_0$ is then equal to fixing the shear stress, $\sigma_0\sim u_0\sqrt{\Delta_\infty} =(u_0^2/u_c)\sqrt{\Delta_c} $. And since the critical shear stress is 
$\sigma_c\sim\sqrt{\Delta_c}\,u_c$, one may rewrite the factor as
\begin{equation}\label{3b-12} 
1-u_0^2/u_c^2=1-\sigma_0/\sigma_c.
\end{equation}
If one chooses to keep $\sigma_s\sim\sqrt\Delta\, u_s$ constant from the
beginning, irrespective how far $\Delta$ has relaxed, one needs to require
$\partial_tu_s=(u_s/2\Delta)\partial_t\Delta$, resulting in a different
proportionality $v_s\sim T_g$ to be inserted into the equations
of motion. The result should be similar.

Comparable calculation and analysis were carried out in~\cite{aging}, using a set of scalar equations 
that may roughly be mapped to the present ones. The quantities $T_g$, $\gamma_1/b\rho$ and 
$\eta_1/b\rho$ (standing for granular temperature, $T_g$-relaxation, and $T_g$-production) were referred to as {\em fluidity, aging parameter,} and {\em rejuvenation parameter}. 
The present consideration is therefore not new, but does provide a treatment embedded 
in {\sc gsh}, hence is transparent and unified,  affording a better founded understanding. 

Since the stress distribution in the experiments of~\cite{aging} is rather nonuniform, there will 
always be some areas with a shear stress close to $\sigma_c$. And the system will tend to cave in there, resulting in a larger strain accumulation than what the average value for $\sigma_s$ 
would predict.   We also note that $Cv_0$, observed to be density-independent in~\cite{aging}, is correlated to the friction angle at  high flow rates [see the discussion above Eq~(\ref{midi})], but postpone a detailed discussion to a future publication.

In the experiment, a very soft spring was used to couple
the fan and the motor. This we believe is an essential reason why this
experiment turned out as observed. Usually, triaxial apparatus with stiff walls are used. And the correcting rates employed by the feedback loop to keep the stress constant are of
hypoplastic magnitudes. As a result, much $T_g$ is excited, and we have the situation of consecutive constant rates, rather than of that of constant stress. The soft spring, as discussed in Sec~\ref{soft springs}, enables quasi-static stress correction without exciting much $T_g$.

\subsubsection{Stability above the Critical Shear Stress\label{scs}}
In the last section, we discussed how an initial temperature $T_0$ and its associated shear rate $v_0$ relax, if the shear stress is hold constant at a value smaller than the critical one, $\sigma_s<\sigma_c$. After the relaxation has completed, the system is in a static and mechanically stable state. The relaxation is slower the closer   $\sigma_s$ is to $\sigma_c$, becoming infinitely slow  for  $\sigma_s=\sigma_c$, or equivalently, $u_0=u_c$. Then we have $\Delta_\infty=\Delta_c$,  $v_s=v_T$, see Eqs~(\ref{3b-7a},\ref{3b-8a}), with especially $T_g\equiv T_0$ remaining a constant, see Eq~(\ref{3b-10}). This is indistinguishable from the rate-controlled, stationary critical state, which we therefore deduce may be maintained at both given rate and  stress. In addition, we note that mechanical stability is lost at $\sigma_s=\sigma_c$ for an elevated $T_g$. In contrast, for $T_g=0$, a granular assembly maintaining a static shear stress exceeding the critical value $\sigma_c$ but not yet breaching the yield condition (as given in Sec.~\ref{yield surfaces}) is stable, though only precariously so: Any $T_g$ sufficiently large, say caused by distant rumbling of the earth, will destabilize it.  

For $u_0>u_c$, or $\sigma_s>\sigma_c$, we have  $A<0$ in Eq~(\ref{3b-10}), and an initial granular temperature $T_0$ will become larger with time, until it diverges. This worsens the instability, and quickens the mechanical collapse, but it does not imply the general impossibility for static stress values larger than $\sigma_c$, because if the initial $T_g$ is too small, of quasi-static values, the factor $h\equiv\sqrt{1+\gamma_o/\gamma_1T_g}$ of Eq~(\ref{2c-5a}) is large, and with it also $u_c=(1-\alpha)/\Lambda\sim h$, see Eq~(\ref{3b-4}). So the factor $1-u_0^2/u_c^2$ remains positive, and a small $T_g$ will relax rather than explode. 

If the initial $T_g$ is large enough, however, it will indeed grow,  and with it also the shear rate $v_s$ -- though they will not diverge, because the system will leave the rate-independent, hypoplastic regime, invalidating the above calculation. The total stress then has a more general form including viscous terms, see Eq~(\ref{sum stress df}), and Eq~(\ref{3b-17}) below. Holding it constant at a value $\sigma_s>\sigma_c$ will again lead to a stationary state, with constant $T_g$ and $v_s$.

Although the shear stress instability  for $\sigma_0>\sigma_c$ holds only for stress-controlled experiments, not rate-controlled ones, the distinction  is not always clear-cut in experiments. For instance, if a step motor is used, and one has a strain versus time curve such as given by Fig~\ref{StepMotor}, than the stress is being hold constant at the plateaus, rendering  the stability of the uniform system fairly precarious even for a strain-controlled approach. This may well be the reason why shear band formation is so frequently observed in the case where the initial density is high and the non-monotonic stress trajectory exceeds $\sigma_c$, see Fig~\ref{fig3}.

We emphasize that this aspect of granular behavior comes out of {\sc gsh} quite naturally, without us ever having put it there. It results from the interplay between yield and the critical state, or more precisely, between the instability of the elastic energy and the stationary solution of the elastic strain. This result does not hinge on any functional dependence of energy or transport coefficients, only on the general  structure of {\sc gsh}.

It is noteworthy that by including a ambient temperature $T_a$, as in Sec~\ref{external perturbation}, 
\begin{equation}
\partial_t T_g=-(\gamma_1/b\rho)h^2[(1-u_0^2/u_c^2)T_g^2-T_a^2],
\end{equation}
see Eq~(\ref{sum T_g}), shifts the equilibrium values  $T_g$ and $v_s$ relax to, from 0 to 
\begin{equation}
T_g\to \frac{T_a}{1-u_0^2/u_c^2},\quad v_s\to \frac{v_T}{u_c^2/u_0^2-1},
\end{equation}
making its effect more pronounced closer to $u_c$, or as discussed above, $\sigma_c$. Note that the second expression is algebraically identical to Eq~(\ref{3b-8}).

\subsubsection{Stress- and Pressure-Control\label{sapc}}

In the last two sections, although the shear stress was controlled, the pressure was not. 
We took $v_{\ell\ell}=0$,  a controlled rate, to keep the density constant and the 
calculation analytical. 
Besides, keeping the shear strain $u_s$ constant is not executable experimentally, and we need to find an experimental prescription that is.
We therefore revisit the situation -- to understand what happens if both the pressure and shear stress are kept constant. Though the general consideration does not appear analytically viable, one simple solution of a 
realistic situation exists: Rewriting Eqs~(\ref{2c-7},\ref{2c-9}) as
\begin{eqnarray}\label{3b-13}
\partial_t\Delta&=&\alpha_1[u_sv_s-u_cv_T(\Delta/\Delta_c)]-(1-\alpha)v_{\ell\ell},
\\\label{3b-14}
\partial_t u_s&=&(1-\alpha)[v_T(u_s/u_c)-v_s],
\end{eqnarray}
and keeping constant $\Delta=\Delta_0$, $u_s=u_0$, we obtain
\begin{equation}\label{3b-15}
v_s=v_T\frac{u_0}{u_c},\quad v_{\ell\ell}=v_T\frac{u_c\alpha_1}{1-\alpha}\left(\frac{u_0^2}{u_c^2}-\frac{\Delta_0}{\Delta_c}\right).
\end{equation}
Taking a $u_0$ such that for given $\Delta_0$ the bracket vanishes, and $v_{\ell\ell}=0$, implies constant density. Inserting $v_s=v_T{u_0}/{u_c}$ into 
the balance equation for $T_g$, Eq~(\ref{sum T_g}), we again obtain Eq~(\ref{3b-9}) 
with (\ref{3b-12}). The only difference is that there is now a clear prescription for the 
experiment, because constant $\Delta,u_s,\rho$ means that pressure and shear stress 
are kept constant. So one may proceed experimentally by applying an arbitrary pressure, then varying the shear stress until the density no longer changes. $T_g$ will then as predicted be long-lived -- infinitely so for $\sigma_s=\sigma_c$, and exploding for  $\sigma_s>\sigma_c$. And no shear stress above $\sigma_c$ is stable if $T_g$ is elevated.


\subsection{Visco-Elastic Behavior } 

All visco-elastic systems such as polymers have a characteristic time $\tau$ that separates two frequency ranges:  fluid-like behavior for $\omega\tau\ll1$, and solid-like one for  $\omega\tau\gg1$.  Like granular media,  polymers are transiently elastic, though the transiency is constant and not variable. The hydrodynamic theory of polymers~\cite{polymer-1,polymer-2,polymer-3,polymer-4}, 
defining an  elastic strain $u_{ij}$, and employing the equation
\begin{equation}\label{poly}
\partial_tu^*_{ij}-v^*_{ij}=-u_{ij}^*/\tau_{ve}, 
\end{equation}
is capable of accounting for many visco-elastic phenomena, including shear-thinning/thicken\-ing, elongational viscosity, the Cox-Merz rule, and the rod-climbing (or Wei\ss enberg) effect. 

As compared to this equation, the granular version, Eq~(\ref{2c-8}), has an extra coefficient 
$\alpha(T_g)$, and its relaxation time varies as $\tau\sim 1/T_g$, with $T_g$ a dynamic variable.  The first difference is not qualitative, as it only accounts for an overall softening. The second difference is crucial, because (1)~the system is truly elastic when the relaxation time diverges for $T_g\to0$, and may sustain a shear stress statically.  (That the static shear stress, as in the case of granular media, has a upper 
limit, is an extra complication not of primary concern here.) (2)~The relation  $1/\tau\sim T_g\sim v_s$ gives rise rate-independence,  while the constancy of $\tau_{ve}$ divides the frequency into solid and liquid regimes.

When there is, in granular media, an ambient temperature $T_a$ much larger than the $T_g$ produced by the imposed shear rate $v_s$, or $T_a\gg T_g$, $v_T\gg v_s$, then polymers and granular media are naturally very similar in their behavior. ($v_T\equiv T_a \sqrt{\gamma_1/\eta_1}$ is the shear rate needed to produce the ambient temperature $T_a$, see Eq~(\ref{v_T}.) The ambient temperature $T_a$ may be maintained by periodic tapping, or is transported by diffusion from another region of great granular activity. The point is,  $T_a$ enables relaxation of the elastic strain and stress for a static system, and implies a vanishing yield stress.

\subsubsection{Creep Motion\label{creep motion}}
In granular media, one frequently observes a shear band, with a boundary between the stationary solid and and the shearing fluid part. More careful experiments (see Komatsu et al~\cite{komatsu}, Crassous et al~\cite{crassous}, and references therein), however, reveals that the transition is not discontinuous, and an exponentially decaying creep motion takes place in the solid. 
To understand this within the framework of {\sc gsh}, we first note that granular temperature being produced in the fluid region will diffuse into the solid one, and is present there 
as an ambient temperature, enabling stress relaxation. This implies a compensating shear rate if the stress is to be maintained. The velocity obtained from integrating the shear rate is the observed creep motion. 

The ambient temperature will decay in space, so will the compensating shear rate. Circumstances are in fact quite similar to that of Sec~\ref{aging}, though 
we need to consider stationary but spatially nonuniform states here: The shear rate needed to compensate stress relaxation is still as given by  Eq~(\ref{3b-7a}); instead of 
Eq~(\ref{3b-8a}), we take $\Delta=\Delta_\infty$; the balance equation for $T_g$ lacks the term $\partial_tT_g$ but contains the diffusive current, see Eq~(\ref{sum T_g}),
\begin{equation}
\nabla^2T_g=(\gamma_1/\kappa_1)[1-u_s^2/u_c^2]\,h^2T_g\equiv T_g/B^2.
\end{equation}
So the decay of $T_g\sim \exp(-x/B)$ is, for an one-dimensional variation along $x$, exponential. Because 
of   Eq~(\ref{3b-7a}), the decay of the shear rate and velocity is also exponential, with the same  
characteristic length $B\sim1/\sqrt{1-u_s^2/u_c^2}=1/\sqrt{1-\sigma_s/\sigma_c}$. That the decay length diverges for $\sigma_s=\sigma_c$ should not surprise, because the solid region ceases to exist then. (Note that $u_c,\sigma_c$ are the critical values at the solid density.)

\subsubsection{Nonlocal Fluidization\label{nonlocal fluidization}}
An striking phenomenon -- discovered by Nichol et al.~\cite{nichol2010}, and  followed up by Reddy et al.~\cite{reddy2011} in a geometry more amenable to  systematic evaluation, is the so-called {\em non-local fluidization} of granular media. In a vessel containing grains, after a shear band is turned on, the medium everywhere, even further away from the band, looses its yield stress, and the Archimedes law is observed to hold: A ball that was stuck at whatever height without the shear band starts to sink or elevate, depending on its density, until it is equal to the granular density. 

The explanation within the framework of {\sc gsh} is: A ball getting stuck in sand builds up an elastic shear stress $\pi_s$ and the associated elastic strain $u_s$ around itself.  Without an ambient temperature, $T_a=0$, this stress holds up the ball's weight if it is not too large, and the ball is stationary. But with one, strain and stress relax, and the ball starts to move -- a striking demonstration of the yield stress vanishing.  

Once the ball is in motion, the elastic strain rate will quickly become stationary in the rest frame of the ball, $\partial_t u_s=0$, reaching a balance between the deformation rate $\sim v_s$ and the relaxation  $\sim T_g u_s$, or $v_s \sim T_g u_s$, see Eq~(\ref{3b-7a}).  Replacing $u_s$ with $v_s/T_g$ in the shear stress $\sigma_s=\pi_s(u_s)$, we turn the elastic stress surrounding the ball into a viscous one, $\pi_s(v_s/T_g)$. If $v_s$ is small enough, we may expand in it, and this viscous stress is linear in the shear flow, $\pi_s\sim v_s/T_g$. Granular media are then Newtonian fluids, with a viscosity $\sim1/T_g$. 

This is the basic {\sc gsh}-explanation, though a word of caution is appropriate here: Any hydrodynamic theory starts from the assumption that the resolution of the theory (``pixel size'') is small compared to system size, but much larger than any micro- and mesoscopic lengths -- in the present case, especially the grain diameter $d$. In~\cite{reddy2011}, the diameter of the probing rod is only $2d$. One may hope that averaging over time and runs will restore the macroscopic limit, but this is far from certain when the two scales are that close.


\subsection{Constitutive Relations}
Granular dynamics is frequently modeled employing  the strategy of
{\em rational mechanics}, by postulating a function $\mathfrak{C}_{ij}$ -- of the  stress $\sigma _{ij}$, strain rate $v_{k\ell}$, and density $\rho $ --  such that the constitutive relation, ${\partial}_{t}\sigma _{ij}=\mathfrak{C}_{ij}(\sigma_{ij}, v_{k\ell}, \rho)$ holds (where ${\partial}_{t}$ is to be replaced by an appropriate objective derivative more generally). It forms, together with the continuity  equation $\partial _{t}\rho +\nabla _{i}\rho v_{i}=0$, momentum conservation, $\partial _{t}(\rho v_{i})+\nabla_{j}(\sigma _{ij}+\rho v_iv_j)=0$, a closed set of equations for $\sigma _{ij}$, the velocity $v_{i}$, and the density $\rho $ (or the void ratio $e$). Both hypoplasticity and barodesy considered below belong to this category. These models  yield, in circumstances where they hold, a realistic account of the complex elasto-plastic motion, providing us with highly condensed and intelligently organized empirical data. This enables us to validate {\sc gsh} and reduce the latitude in specifying  the energy and transport coefficients.  

At the same time, one needs to be aware of their drawbacks, especially the more hidden ones. First of all is the apparent freedom in fixing $\mathfrak C$ -- constrained only by the data one considers, not by energy conservation or entropy production that were crucial in deriving {\sc gsh}.  This is what we believe the main reason why there are so many competing engineering models. Worse, this liberty explodes when one includes gradient terms. So most models refrain from the attempt to account for nonuniform situations, say elastic waves. Second, in dispensing with the variables $T_g$ and $u_{ij}$, and choosing the shortcut via $\mathfrak C(\sigma_{ij}, v_{k\ell}, \rho)$, one reduces the model's range of validity and looses the benefit of $u_{ij}$'s simple behavior: First, the models of hypoplasticity and barodesy are valid only for  $T_g\sim|v_s|$, so a $T_g$ that is either too small or oscillates too fast will invalidate these models, as will a $T_g$ derived from an external source, such as considered in  Sec~\ref{external perturbation}.  Second, as the analytical solution of the approach to the critical state in Sec~\ref{approach critical} shows, considering $u_{ij}$ -- though it is not directly measurable -- is a highly simplifying intermediate step. The case for  $u_{ij}$ is even stronger, when considering proportional paths and the barodesy model.  

\subsubsection{The Hypoplastic Model\label{Hypoplasticity}} 

The {\em hypoplastic model} starts from the postulated, rate-independent
constitutive relation, 
\begin{equation}\label{3b-1}
\partial_t\sigma_{ij}=H_{ijk\ell}v_{k\ell}+
\Lambda_{ij}\sqrt{v_s^2+\epsilon v_{\ell\ell}^2}, \end{equation} where
$H_{ijk\ell},\Lambda_{ij},\epsilon$ are (fairly involved) functions of the
stress and packing fraction~\cite{kolymbas1}. Incremental nonlinearity as discussed in Sec~\ref{Load and Unload} is also part of the postulate. The simulated granular response is realistic for deformations at constant rates.
 
{\sc gsh} reduces to the hypoplastic model in the hypoplastic regime, for $T_g\sim|v_s|$, $\alpha =\bar \alpha,\alpha_1=\bar\alpha_1$,  $P_T, \eta_1T_g
v^0_{ij}\to0$. This is because $\sigma_{ij}=(1-\bar\alpha)\pi_{ij}$ of Eq~(\ref{sum stress hp}) is then, same as $\pi_{ij}$, a function of $u_{ij}, \rho$,  and we may write $\partial_t\sigma_{mn}=({\partial\sigma_{mn}}/{\partial u_{ij}})\partial_tu_{ij}+
({\partial\sigma_{mn}}/{\partial\rho})\partial_t\rho$.
Replacing $\partial_t\rho$ with the first of Eq~(\ref{2c-1}),
$\partial_tu_{ij}$ with Eq~(\ref{2c-8}), using Eq~(\ref{TgVs}) to eliminate $T_g$, we arrive at an equation with the same structure
as Eq~(\ref{3b-1}). Our derived result for $H_{ijk\ell},\Lambda_{ij}$ is
different from the postulated engineering expressions, and somewhat simpler, but they yield very
similar {\em response ellipses}, see~\cite{granL3}. (Response ellipses are
the strain increments as the response of the system, given unit stress
increments in all directions starting from an arbitrary point in the stress
space, or vice versa, stress increments as the response for unit strain increments.)

\subsubsection{Proportional Paths and Barodesy}

Barodesy is a very recent model, again proposed by Kolymbas~\cite{barodesy}. As compared to hypoplasticity, it is more modular and better organized, with different parts in $\mathfrak C_{ij}$ taking care of specific aspects of granular deformation, especially that of {\em proportional paths}. We take {\sc p}$\varepsilon${\sc p} and {\sc p}$\sigma${\sc p} to denote, respectively, proportional strain and stress path.  Their behavior is summed up by the Goldscheider rule ({\sc gr}): (1)~A {\sc p}$\varepsilon${\sc p} starting from the stress $\sigma_{ij}=0$ is associated with a  {\sc p}$\sigma${\sc p}.
(2)~A {\sc p}$\varepsilon${\sc p} starting from $\sigma_{ij}\not=0$
leads asymptotically to the corresponding  {\sc p}$\sigma${\sc p} obtained when starting at $\sigma_{ij}=0$.
(The initial value $\sigma_{ij}=0$ is a mathematical idealization, neither easily realized nor part of the empirical data that went into {\sc gr}. We take it  {\em cum grano salis}.) 

Explanation: Any constant strain rate $v_{ij}$ is a {\sc p}$\varepsilon${\sc p}. In the principal strain axes  $(\varepsilon_1,\varepsilon_2,\varepsilon_3)$, a constant $v_{ij}$ means the system moves with a constant rate along its direction, with $\varepsilon_1/\varepsilon_2=v_1/v_2,\, \varepsilon_2/\varepsilon_3=v_2/v_3$ independent of time. What {\sc gr} states is that there exists an associated stress path that is also proportional, also a straight line in the principal stress space, that there are pairs of strain and stress path which are linked,  and if the initial stress value is not on the right line, it will converge onto it.  

If {\sc gsh} is as claimed a broad-ranged theory on granular behavior, we should be able to  understand {\sc gr}  with it, which is indeed the case. Given any constant rate $v_{ij}$, the elastic strain will -- irrespective of its initial value, relax into the stationary state of Eqs~(\ref{2c-7},\ref{2c-9}),
\begin{equation}
\label{eq74}
u_c=\frac{1-\alpha}{\Lambda},\quad \frac{\Delta_c}{u_c}=\frac{\alpha _1}{\Lambda _{1}}+\frac{1-\alpha}{u_c\Lambda_1}\frac{v_{\ell\ell}}{v_s},
\end{equation}
with ${u_{ij}^{\ast }|_c}/{u_c}={v_{ij}^{\ast }}/{v_s}$.  Adding in the information from  Eqs~(\ref{2b-2a},\ref{2b-2b}), we also find
\begin{equation}
{\sigma_{ij}^{\ast }}/{\sigma_s}(\rho)={v_{ij}^{\ast }}/{v_s}.
\end{equation}
If the strain path is isochoric, with $v_{\ell\ell}=0$ and $\rho=$ const, both the deviatoric strain and stress are dots that remain stationary and do not walk down a path as time progresses. Clearly, these are simply the ideally plastic, stationary, critical state that we considered in Sec~\ref{critical state}. If $v_{\ell\ell}\not=0$ with the density $\rho[t]$ changing accordingly, ${u_{ij}^{\ast }|_c}$ and ${\sigma_{ij}^{\ast }}$ will walk down a straight line along ${v_{ij}^{\ast }}/{v_s}$, with a velocity determined, respectively, by $u_c(\rho[t])$ and $\sigma_s(\rho[t])$. 

Given an initial strain deviating from that prescribed by Eq~(\ref{eq74}),  $u_0\not=u_c,\Delta_0\not=\Delta_c$, Eqs~(\ref{2c-7},\ref{2c-9}) clearly state that the deviation will exponentially relax, until they vanish -- ie., the strain and the associated stress will converge onto the prescribed line. All this is very well, but {\sc gr} states that it is the total stress that walks down a straight line. With
\begin{equation}
\pi_{ij}=P_\Delta(\rho)[\delta_{ij}+(\pi_s/P_\Delta)v_{ij}^*/v_s],
\end{equation}
this fact clearly hinges on $(\pi_s/P_\Delta)$ -- a function of $\Delta/u_s$ [see Eq~(\ref{2b-3b})] --  not depending on the density. As long as $v_{\ell\ell}\ll v_s$, we have  $\Delta_c/u_c\approx{\alpha _1}/{\Lambda _{1}}$, a combination that we did assumed is density independent, see Eq~(\ref{density-dependence}), partially in anticipation of the fact that the friction angle of the critical state, a function of $(\pi_s/P)$, is independent of the density. And $v_{\ell\ell}/v_s$ must indeed remain small to avoid hitting either $\rho_{cp}$ or $\rho_{lp}$ too quickly.

In~\cite{GSH&Barodesy}, the results of {\sc gsh} are compared to that of barodesy, with mostly quantitative agreement. (The energy of Eq~(\ref{2b-6}) was employed there. So the results are more realistic.) When looking at $\mathfrak C_{ij}$, it is easy to grasp that  the construction of a constitutive relation is only for someone with vast experience about granular media. That we could substitute this deep knowledge with the notions of {\em variable transient elasticity}, giving rise to a theory just as capable of accounting for elasto-plastic motion, is eye-opening. It suggests that sand, in its qualitative behavior, may be after all neither overly complicated, nor such a rebel against general principles.


\subsection{Elastic Waves\label{elastic waves}}
That elastic waves propagate in granular media~\cite{jia1999,jia2004} is an important fact, because it is an unambiguous proof that granular media possess an 
elastic regime, and behave as elastic media in certain parameter ranges. Experimental exploration of the elastic to plastic transition would be equally crucial, and elastic waves remain a useful tool for this purpose.  

There is a wide-spread believe that small, quasi-static increments from any 
equilibrium stress state is elastic, but large ones are plastic.  As discussed in 
Sec~\ref{3regimes}, this assumption is illogical, because a large increment is 
the sum of small ones. In {\sc gsh}, the parameter that sets the boundary between elastic and plastic regime is the granular temperature $T_g$. We have 
quasi-elastic regime for vanishing $T_g\sim v_s^2$, and the 
hypoplastic one for elevated $T_g\sim v_s$. 

A perturbation in the elastic strain or stress propagate as a wave only in the quais-elastic regime, while it diffuses in the hypoplastic one. 
More specifically, we shall derive a telegraph equation from {\sc gsh}, with a quantity $\sim T_g$ taking on the role of the electric resistance~\cite{zhang2012}. It defines a characteristic  frequency $\omega_0\sim T_g$, such that elastic perturbations of the frequency $\omega$ diffuse for $\omega\ll\omega_0$, and propagate for $\omega\gg\omega_0$.  In the quasi-elastic regime, $\omega_0\to0$, and all perturbations propagate. In the hypoplastic regime, when $T_g$ gets elevated, so does $\omega_0$, pushing the propagating range to ever higher frequencies. Eventually, the associated wave length become comparable to the granular diameter,  exceeding  {\sc gsh}'s range of validity.

To derive the telegraph equation, we start with two basic equations of {\sc gsh}, Eqs~(\ref{2c-8},\ref{2c-1}),
\begin{eqnarray}  \label{ew1}
\rho\partial_tv_i-(1-\alpha)\nabla_mK_{imkl}u^*_{kl}=0, \\
\partial_tu^*_{ij}-(1-\alpha)v^0_{ij}=-\lambda T_gu^*_{ij},  \label{ew2}
\end{eqnarray}
where $K_{imkl}\equiv-\partial^2w/\partial u_{im}\partial u_{kl}$. (For simplicity, we concentrate on shear waves, assuming $v_{\ell\ell}\equiv0 $.) 
For $T_g\to0$, both plastic terms $\lambda T_gu^*_{ij}$ and $\alpha\sim T_g$ are  negligibly small, such that these two equations reproduce conventional elasticity theory. The variation of wave velocities $c$ with stress and density is then easily calculated, because $c^2$ is given by the eigenvalues of the matrix $K_{imnj}q_mq_n/(\rho q^2)$  ($q_m$ is the wave vector). The result~\cite{ge4} agrees well with observations~\cite{jia2009}.

There are two ways to crank up $T_g$. First is to introduce an ambient temperature, such as by tapping or a remote shear band, second is to increase the amplitude of the wave mode, because its own shear rate also creates $T_g$. The granular temperature has a characteristic time $\tau_T=b\rho/\gamma_1$, see Eq~(\ref{sum T_g}), that is of order $10^{-3}$~s in dense media. For simplicity, we assume that the  wave mode's  frequency is much larger than $1/\tau_T$ , such that $T_g$ and $\alpha(T_g)$ are essentially constant. This implies
\begin{eqnarray}\label{ew3}
(\partial^2_t+\lambda T_g\partial_t)\,u^*_{ij}={\textstyle\frac12}(1-\alpha)^2\times
\\
\nabla_m[K_{imkl}\nabla_ju^*_{kl}+K_{jmkl}\nabla_iu^*_{kl}].
\nonumber
\end{eqnarray}
Concentrating on one wave mode propagating along $x$, with $c_{\rm qs}$ the quasi-elastic velocity and $\bar u\sim e^{iqx-i\omega t}$ the amplitude of the associated eigenvector, we obtain the telegraph equation, 
\begin{equation}\label{ew4}
(\partial^2_t+\lambda T_g\partial_t)\,\bar u= (1-\alpha)^2c_{\rm qs}^2\nabla_x^2\,\bar u
\equiv c^2\nabla_x^2\,\bar u.
\end{equation}
(The coefficient $\alpha$ accounts for the fact that granular contacts soften with $T_g$, and the effective elastic stiffness decreases by $(1-\alpha)^2$. In the language of electromagnetism, 
 $(1-\alpha)^{-2}$  is a dielectric permeability.)
Inserting $\bar u\sim e^{iqx-i\omega t}$ into Eq~(\ref{ew4}), we find
\begin{equation}
c^2 q^2={\omega^2+i\omega\lambda T_g},
\end{equation}
implying diffusion for the low frequency limit, $\omega\ll\lambda T_g$,
\begin{equation}
q\approx\pm\frac{\sqrt{\omega\lambda T_g}}c \, \frac{1+i}{\sqrt2},
\end{equation}
and propagation for the high-frequency limit, $\omega\gg\lambda T_g$,
\begin{eqnarray}
q&\approx&\pm\frac\omega c\left(1+i\,\frac{\lambda T_g}{2\omega}
\right),
\\
\bar u&\sim&\exp{\left[-i\omega\left(t\mp\frac xc\right) 
t\mp x\frac{\lambda T_g}{2 c}\right]}.
\end{eqnarray}
The first term in the square bracket accounts for wave propagation, the second a decay length $2c/\lambda T_g$, which is frequency-independent  if $T_g$ is an ambient temperature. If  $T_g$ is produced by the elastic wave itself,  it varies between $T_g\sim v_s^2\sim\omega^2q^2\sim\omega^4$ and $T_g\sim v_s\sim\omega^2$ depending on the amplitude, and the decay length is strongly frequency dependent.
     
A brief wave pulse, arbitrarily strong, can always propagate through granular media if its duration is too brief to excite sufficient $T_g$ for the system to enter the hypoplastic regime. The duration must be much smaller than the characteristic time $b\rho/\gamma$ of $T_g$, see Eq~(\ref{sum T_g}).

\section{Rapid Dense Flow\label{dense flow}}
In Chapter~\ref{hypoplastic motion} on hypoplastic motion, the seismic pressure $P_T$ and the viscous shear stress $\sim\eta_g$ were neglected. In this chapter, we consider flows in which they are important, even dominant. Including them, we are leaving the rate-independent, hypoplastic regime. Being  quadratic in the shear rate, the correction come on slowly. This may be the reason rate-independence was widely perceived as a basic property of granular media in soil mechanics.

\subsection{Uniform Dense Flow\label{udf}} 
\subsubsection{Constant Density Experiments}
Starting from Eq~(\ref{sum stress df}), we first substitute the unspecific elastic contributions $(1-\bar\alpha)P_\Delta$,  $(1-\bar\alpha)\pi_s$ with their critical state expressions,  Eqs~(\ref{3b-4a}), because these are the steady state values the elastic strain will acquire for constant shear rates, 
\begin{equation}\label{3b-17}
P=P_c+\frac{T_g^2}2\frac{a\rho^2b}{\rho_{cp}-\rho},
\quad\sigma_{s}=\sigma_c+\eta_1T_gv_s.
\end{equation}
%
For a stationary temperature, $T_g=|v_s|\sqrt{\eta_1/\gamma_1}$ [see Eq~(\ref{TgVs})], we may abbreviate them as 
\begin{equation}\label{3b-17a}
P=P_c+e_1v_s^2,\quad \sigma_s=\sigma_c+e_2v_s^2,
\end{equation} 
noting that $P_c,\sigma_c,e_1,e_2$, being functions of the density are constant if the density is. 
Taking 
\begin{equation}\label{3b-18}
\frac{\sigma_s}{P}=\frac{\sigma_c+e_2v_s^2}{P_c+e_1v_s^2}
\end{equation}
as a friction angle, we have a change between two constant values, from $\sigma_c/P_c$ for 
$v_s\to0$, to $e_2/e_1$ for $v_s\to\infty$. Whether the change occurs with $v_s$, or slightly 
more quickly with $v_s^2$, is here of a fairly subtle difference -- though it is of course not in 
Eq~(\ref{3b-17a}): That pressure and shear stress grow as $v_s^2$ in the fast limit was already observed by Bagnold~\cite{Bagnold}. Note that since no elastic solution is stable for $\rho<\rho_{\ell p}$, we set $P_c,\sigma_c=0$ for any density below  $\rho_{\ell p}$, where the system is always in the Bagnold regime, $P,\sigma_s\sim v_s^2$. 

Both $\eta_1,\gamma_1$ are believed~\cite{Bocquet} to diverge for $\rho\to\rho_{cp}$. 
Taking $\eta_1\sim(\rho_{cp}-\rho)^{-1.5}$,  $\gamma_1\sim(\rho_{cp}-\rho)^{-0.5}$, or $e_1,e_2\sim(\rho_{cp}-\rho)^{-2}$, implies first the independence of the high rate friction angle, $e_1/e_2$, from $(\rho_{cp}-\rho)$, and  second,
\begin{equation}\label{midi}
P-P_c\sim v_s^2/(\rho_{cp}-\rho)^{2}.
\end{equation}•
There are some experimental evidences for both~\cite{denseFlow}, though the data appear  different for glass and polystyrene beads~\cite{Savage}.

\subsubsection{Shear Stress Minimum\label{stress minimum}}

If the shear experiment is not executed at given density, but rather at given pressure $P=P_0$, circumstances are more complicated. First, the rate dependence of the friction angle is the same as that of the shear stress alone,
\begin{equation}\label{3b-18a}
{\sigma_s}/{P_0}={(\sigma_c+e_2v_s^2)}/P_0.
\end{equation}
Second, crucially, the density is a function of the rate: Inverting the first of Eq~(\ref{3b-17a}) and defining $v_s^2=(P_0-P_c)/e_1\equiv f(\rho)$, we have 
$\rho=f^{-1}(v_s^2)$. As a result, the rate dependence in pressure controlled experiments  
hinges on the density dependence of the transport coefficients, which combine to form $P_c$ and $e_1$. 

Third, $\sigma_s$ and the friction angle are, as observed in~\cite{Brodsky1,Brodsky2}, no longer necessarily monotonous functions of the shear rate $v_s$. In this context, it is important 
to realize that in a nonuniform geometry, keeping the volume constant does not usually maintain a constant density. So a non-monotonic relation between $\sigma_s$ and $v_s^2$ may also happen for constant volume, especially if the shear rate is strongly nonuniform. 

To better understand the last point, consider two uniform subvolumes (instead of the continuous non-uniformity of the experiments). They are 
in contact via a flexible membrane, such that
their total volume  $V_1+V_2$ is a constant. Initially, the
total system is uniform, with both densities equal,
$\rho_1=\rho_2$, and both shear rates vanishing,
$\dot\gamma_1,\dot\gamma_2=0$. Now, if $\dot\gamma_2$ is
cranked up, but $\dot\gamma_1$ remains zero, because 
$P_1(\rho_1,\dot\gamma_1)=P_2(\rho_2,\dot\gamma_2)$, the
density must change and the membrane will stretch in one
direction, typically with $\rho_2$ decreasing and
$\rho_1$ increasing. If system~1 is much larger than 2,
the stretching of the membrane will not change $\rho_1$
much, as a result, $P_1(\rho_1,\dot\gamma_1)$ will remain
essentially constant. So will $P_2=P_1$. As a result,
the pressure-controlled limit holds. More
realistically, if both systems are comparable in size, an
intermediate case between the pressure- and
density-controlled limit will take place. As only in the
strictly density-controlled limit do we have
monotonicity of the shear stress, any inhomogeneity in the 
shear rate may result in non-monotonic behavior of the
shear stress.

\subsubsection{Comparison to Other Models}
First, we compare {\sc gsh} to the continuum theory that Boquet et al.~\cite{Bocquet} developed to account for their experiment, see also~\cite{luding2009}. The theory includes the Cauchy stress $\sigma_{ij}$, and a balance equation for the temperature $T_G\sim T_g^2$ [see Eq~(\ref{2b-2c})]. For $v_{\ell\ell}=0$, they are:
\begin{eqnarray}\label{3b-19}
\sigma_{ij}&=&P\delta_{ij}-\eta v^0_{ij},\\\nonumber \partial_t
T_G&\sim&\eta v_s^2-\gamma \, T_G +\nabla_i(\kappa\nabla_iT_G),
\\\nonumber
\text{with}&&
P\sim{T_G}, \quad \eta,
\gamma, \kappa\sim
{\sqrt{T_G}}.
\end{eqnarray}
Comparing these to the above dense flow expressions of Eqs~(\ref{sum T_g},\ref{sum stress df}), we find agreement except for the fact that the elastic contributions $P_c,\sigma_c$ are missing.  

\begin{figure}[tbh] \begin{center}
\includegraphics[scale=0.5]{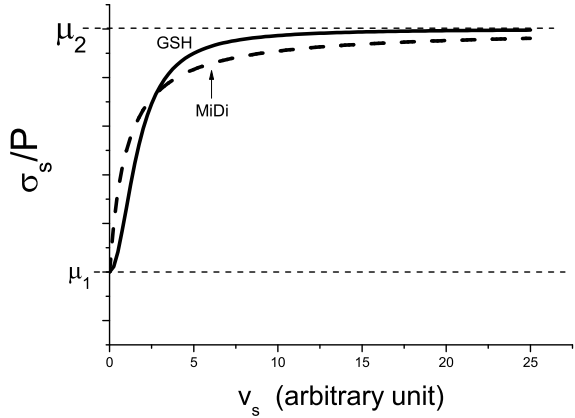}
\end{center}
\caption{\label{MiDi}
Comparison of the MiDi constitutive relations with GSH.}
\end{figure}

Next, we compare {\sc gsh} to the MiDi constitutive relations.  Starting from the postulate that granular rheology in dense flows is
controlled by the dimensionless parameter of inertial number,
$I\sim\dot\gamma/\sqrt P$, Pouliquen et al. distilled two
locally applicable constitutive relations from
experiments and simulations, for the density and the friction angle $\sigma_s/P$, see~\cite{pouliquen1,pouliquen2},
\begin{eqnarray}
\label{3b-20} \rho_{cp}-\rho\sim I, \quad
\frac{\sigma_s}{P}=\frac{\mu_1+\mu_2\,I}{1+I}.
\end{eqnarray}
Identifying $\dot\gamma\to v_s$, the first relation may be combined to form $P\sim v_s^2/(\rho_{cp}-\rho)^2$, same as Eq~(\ref{midi}) if $P_c$ is neglected. 
This is to be expected, because  the inertial number $I$ (as Savage observed~\cite{Savage})
does not contain any elastic  information.  The second expression is similar to Eq~(\ref{3b-18}), and  it does contain elastic contributions, because we may identify $\mu_1=\sigma_c/P_c$, and 
$\mu_2=e_2/e_1$. The transition between the two friction angles is linear in the reduced shear rate $I$, not quadratic as in {\sc gsh}, though this is as mentioned a subtle difference, see Fig~\ref{MiDi}. 

More troubling is the claim that Eqs~(\ref{3b-20}) are valid for pressure controlled experiments, not volume controlled ones -- implying that the friction angle $\sigma_s/P_0$ tends to a constant for large shear rates. In contrast, {\sc gsh} contends that it behaves as $\sigma_s/P_0\to e_2(\rho)v_s^2/P_0$, see Eq~(\ref{3b-18a}). It is generally a constant only if the rate 
dependence of $e_2(\rho)$ cancels that of $v_s^2$ -- universally. As mentioned above, density dependence of transport coefficients do vary. So we expect the friction angle to display a more diverse behavior, and may even diverge. 

On the other hand, the majority of experiments cited in~\cite{midi} are inhomogeneous, with 
varying shear rate and density, so controlling the volume does not mean holding the local density constant. In princinple, of course,  {\sc gsh}  is capable of dealing with these situations --  though before ons can solve this set of partial differential equations for given boundary conditions, we need to have clarified the density dependence of all transport coefficients. 

\subsection{Jamming, or Angle of Repose\label{jamming}}
Considering shallow flows on an inclined plane and rotating drums, Aranson and Tsimring identified the
hysteresis of transition, or the delay between jamming and fluidization, as a key feature of granular behavior~\cite{aranson}. 
Their theory takes the stress $\sigma_{ij}$ as the sum of two parts, one solid-like, $\hat\varrho\pi_{ij}$, and a rate-dependent fluid one. $\hat\varrho$ is an 
order parameter that is 1 for granular solid, and 0 for dense flow. The authors take the friction angle $\varphi$, differently than above, as the ratio of the solid stress components, rather than the total ones, and postulate a 
free energy $f(\hat\varrho)$ such that $\hat\varrho=1$ is unstable for large  friction angles, $\varphi> \varphi_{S}$; while $\hat\varrho=0$ is unstable for a small ones, $\varphi<\varphi_{R}$. But both are stable in the intermediate region, $\varphi_{S}>\varphi>\varphi_{R}$. ($\varphi_{S}$ and $\varphi_{R}$ are referred to as the angle of stability and repose, respectively.) Though this theory does not consider variations in $\rho$ or $T_g$, and takes the solid stress $\pi_{ij}$ as an input from some other theory, it provides a pivotal insight: The viability, even appropriateness, of using a partially bistable energy to account for the hysteresis.

Turning now to {\sc gsh}, we first note that for rate-controlled experiments in the hypoplastic regime, there is no hysteresis, only uniform and continuous paths to the ideally plastic, critical state and back, all given by Eqs~(\ref{3b-17}), with the terms $\sim v_s^2$ evident only at higher rates. The second type of fluidization  takes place either at quasi-static rates or stress-controlled. It is discontinuous, and happens because a yield surface is breached -- say when the ratio $\pi_s/P_{\Delta}$ is larger than $\sqrt{2{\cal A}/{\cal B}}$, see~Eq~(\ref{2b-3}).  This is what we consider now. 

On a plane inclined by a slowly increasing angle $\varphi$,  with $y$ denoting the depth of the granular layer on the plane, and $x$ along the slope, we take the stress to be $\sigma_{xx}=\sigma_{yy}=\sigma_{zz}=P$, $\sqrt2\,\sigma_{xy}=\sigma_s$, $\sigma_{yz}=\sigma_{xz}=0$. Integrating $\nabla_j\sigma_{ij}=g_i\rho$ assuming spacial dependence only along $y$, we find $\sigma_{xy}=g\sin\varphi\int\rho(y)dy$,  $\sigma_{yy}=g\cos\varphi\int\rho(y)dy$, or $\tan\varphi=\sigma_{xy}/\sigma_{yy}=\sigma_s/\sqrt2\,P=\pi_s/\sqrt2\, P_{\Delta}\leq\sqrt{{\cal A}/{\cal B}}$, implying a stability angle $\varphi_{S}$ given by
\begin{equation}\label{sb12}
\tan\varphi_{S}=\sqrt{{\cal A}/{\cal B}}.
\end{equation}

Jamming, the reverse transition, is a drop of the shear rate $v_s$  to zero, at given shear stress and elevated $T_g$, which therefore takes place as in Sec~\ref{aging}. And because only stress values smaller than $\sigma_c$ will jam, and come to a stand still, the angle of repose $\varphi_{R}$ is 
\begin{equation}
\tan\varphi_{R}={\sigma_c}/\sqrt2P_c.
\end{equation}
On a plane inclined by a slowly decreasing angle $\varphi$, the seismic and viscous terms $\sim v_s^2$ are small in the vicinity of $\varphi_{R}$, and were neglected. Note both ${\sigma_c}(\rho), P_c(\rho)$ are functions of $\rho$, the value of which varying with $y$ is not always fixed. But the ratio ${\sigma_c}/P_c$ (a function of ${\Delta_c}/{u_c}={\alpha_1}/{\Lambda_1}$) is density independent, see the discussion below Eq~(\ref{3b-5}).

The difference between the two angles is clearly a question of $T_g$, with $T_g\to0$ in the first, and elevated in the second, case.  We have $\varphi_{R}<\varphi_{S}$,
because the critical state is only realized and defined, if the yield surface is not breached in an approach to the critical state. 

\section{Compaction\label{compaction}}
Compaction -- a slow increase of the density at given pressure when the system is perturbed -- is a ubiquitous phenomenon in granular media, though not a universal one. For instance, the density is observed to both in- and decrease approaching the critical state. Within the framework of {\sc gsh}, compaction stems from the basic fact that the elastic compression $\Delta$  relaxes in the presence of $T_g$. Keeping the pressure constant, the density must increase to compensate for the diminishing $\Delta$. This is also the case approaching the critical state, though with the difference that, in addition to the relaxation that reduces $\Delta$, the applied shear rate $v_s$ increases it -- consider Eq~(\ref{2c-7}), or $\partial_t\Delta -\alpha_1u_sv_s =-\lambda_1T_g\Delta$. In approaching the critical state, $v_s\sim T_g\not=0$, $\Delta$ may in- or decrease, depending how large $u_s,\Delta$ are at any moment, see Sec~\ref{pressure approach}.

\subsection{Reversible and Irreversible Compaction}
Consider the pressure $P=P_\Delta+P_T$ assuming vanishing shear rate, $v_s=0$, with $P_\Delta$ the elastic, and $P_T$ the seismic, contribution, see Eqs~(\ref{2b-2b},\ref{2b-4},\ref{3b-17}),
\begin{eqnarray}\label{gc5}
P_\Delta&=&{(1-\alpha)}{\cal B}(\rho)\Delta^{1.5},\\P_T&=&{T_g^2}({a\rho^2 b})/{2(\rho_{cp}-\rho)},\nonumber
\end{eqnarray}
where both $\cal B$ and $P_T$ are monotonically increasing functions of $\rho$. So the density must get larger when $\Delta$ decreases. It is irreversible because the relaxation of $\Delta$ is. 

At small $T_g$, the relaxation of $\Delta$ is slow, and the seismic pressure $P_T$ may be neglected.  
This is the limit most soil mechanical experiments are in. Here, only irreversible compaction is 
observed. For larger $T_g$, the seismic pressure must be included. Then the relaxation of $\Delta$ 
for given $T_g$ increases $P_T$ and decreases $P_\Delta$ , such that $P=P_\Delta+P_T$ remains 
constant. After the relaxation of $\Delta $ has run its course, with $P_\Delta\to0$, if one modifies 
$T_g$ but maintains $P=P_T$, the density will change in response, in both directions and hence 
reversibly. Note this is where the fluid equilibrium condition Eq~(\ref{2a-2}), obtained by maximizing 
the true entropy, holds. Consequently, the relaxation of $\Delta$ occurs because it increases the entropy.

\subsection{History Dependence versus Hidden Variable\label{hdvhv}}

Changing $T_g$ midway at constant $P$, with $\Delta$ still finite,
will mainly lead to a change in $\Delta$,  because the density responds much more slowly. 
It disrupts the relaxation of $\Delta$, in essence resetting its initial
condition. This phenomenon was observed in~\cite{mem} and interpreted as a
memory effect. Generally speaking, ``memory" is
usually a result of hidden variables: When the system
behaves differently in two cases, although all state
variables appear to have the same values, we speak of memory-, or
history-dependence. But an overlooked variable that
has different values for the two cases will naturally
explain the difference. In the case of compaction, the manifest and hidden
variables are  $\rho$ and $\Delta$, respectively.

\subsection{Tapping and the Edward Entropy\label{tapping}}

Numerous experiments have shown that tapping leads to irreversible compaction and reversible 
density change, see the review article~\cite{1nico}. It is usually accounted for by the 
specifically tailored {\it granular statistical mechanics}~\cite{Edw} and the Edward 
entropy $S_{Ed}$, or some variant of it. Substituting the volume
$V$ for the energy $E$, and compactivity $X$ for the
temperature $T$, this theory employs ${\rm d}V=X{\rm
d}S_{Ed}$ as the basic thermodynamic relation for a
{\it``mechanically stable agglomerate of infinitely
rigid grains at rest"}~\cite{Edw}. The entropy
$S_{Ed}$ is obtained by counting the number of
possibilities to package grains stably for a given volume,
equating it to $e^{S_{Ed}}$. 
Compaction is taken as an
indication of an increasing  $S_{Ed}$.

Two reasons prompt us to doubt  its appropriateness. 
First, the number of possibilities to arrange
grains concerns inter-granular degrees of freedom. These are vastly overwhelmed by the much more numerous configurations of the inner-granular degrees of freedom. In other words, the Edward entropy $S_{Ed}$ is a special case of the granular entropy $S_g$, and as discussed in Sec~\ref{intro-3}, we always have $S_g\ll S$. {\em In equilibrium, where Eq~(\ref{2a-2}) holds, the entropy $S$ is maximal, and macroscopic energy minimal.} This is unrelated to the number of possibilities to package grains. 
One would be able to neglect $S$ and concentrate on $S_g$ if these two were only weakly connected, if the energy decay from $S_g$ to $S$ were exceedingly slow. This is not the case. The relaxation of $s_g$ or $T_g$, via inelastic scattering, is a fast process.

Second, even assuming a weak coupling between $S$ and $S_g$,   the Edward entropy
$S_{Ed}$ would, as defined, still not be a relevant measure: The actual starting point of the Edward entropy is the assumption that  $S_g$ does not depend on the energy $E$, which is always zero for infinitely rigid,  non-interacting grains at rest, however they are packaged. Taking the entropy generally as a function of energy and volume, $S_g(E,V)$, we have, quite generally,  
\[{\rm d}S_g=\frac{\partial S_g}{\partial E}{\rm d}E+\frac{\partial
S_g}{\partial V}{\rm d}V \equiv\frac1{T_g}{\rm d}E+\frac P{T_g}{\rm d}V.\] 
Usually, one keeps the volume constant, ${\rm d}V=0$, and consider the conventional 
expression, ${\rm d}S_g=(1/T){\rm d}E$. Taking instead $E\equiv0$, we have ${\rm d}S_g=(P/T)
{\rm d}V$, equivalent to the Edward expression ${\rm d}V=(T/P){\rm d}S_g\equiv X{\rm d}S_g$. 

This construction ignores three essential points: First, perturbing the system, allowing it to explore the phase space, introduces kinetic energy that one must include. Then clearly, $E\not\equiv0$. Second, because of the Hertz-like contact 
between grains, little material is deformed at first contact, and the compressibility diverges at
vanishing compression. This is a geometric fact independent of how rigid the bulk material is. 
Therefore, infinite rigidity is never a realistic limit in granular media, and there is always considerable 
elastic energy stored among grains in mechanically stable agglomerates -- even at finite perturbation, 
as long as $\Delta$ is not zero.  Finally, $S_{Ed}$ as defined is the granular entropy at vanishing granular motion and compression. Its phase space is therefore severely constrained. Generally speaking, each classical particle has states in a 6D-space, three for positions and three for the velocities. $\exp(S)$ is the number of states times the Loschmidt's number; $\exp(S_g)$ is the number of states times the number of grains, and $\exp(S_{Ed})$ is the number of states in 3D space (no velocities)  times the number of grains. Therefore  
\begin{equation}
S_{Ed}\ll S_g\ll S.
\end{equation}
Going toward equilibrium, a system searches for the greatest number of states to equally redistribute its energy. One bears the burden of proof for the claim that it is sensible for the system to neglect $S$ and concentrate on $S_{Ed}$. In contrast, {\sc gsh} identifies compaction as a process taking place at finite $T_g$ and compares the true entropy $S$ of macrostates at that $T_g$. It also accounts for entropy increase, by detailing how macroscopic energy decays into granular heat, and how this is converted to true heat.

Reversible and irreversible compaction as accounted for by {\sc gsh} is a universal granular 
phenomenon. It occurs at given pressure and  $T_g$, however $T_g$ is created. At the same time, numerous experiments show that tapping, though especially efficient,  is but one way to achieve compaction, leading to results similar to that of other methods~\cite{1nico}.  So it is natural to take the consideration of the last section to hold for tapping as well. This should be true for gentle tapping, but the connection to stronger ones warrants further scrutiny.  

Gentle tapping leads to granular jiggling and a small $T_g$,  
though one that fluctuates in time, with periodic flare-ups. As long as  $P_T$ may be neglected, 
$\Delta$ will relax according to the momentary value of $T_g$, haltingly but monotonically. 
Since the relaxation is a slow process, one could average over many taps to yield a coarse-grained 
account. Given a granular column with a free upper surface in the 
gravitational field, because a given layer is subject to a constant pressure, the density will increase to  compensate for the diminishing $\Delta$. The characteristic time of $\Delta$-relaxation diverges towards the end, and is not a constant, see~\cite{compaction}.  

Stronger tapping leads to a higher $T_g$, with $\Delta$ relaxing more quickly.  
$P_T$ must now be included. Periodically, when all grains are at rest, $P_T$ vanishes, and  $\Delta$ is necessarily increased to maintain the given pressure. This introduces a non-monotonicity into $\Delta(t)$, and raises the question, whether the system, when again at an elevated $T_g$, will pick up the relaxation of $\Delta$ where it was left when the system last crushed to a stop. And why it should do so. If it does, we can again take tapping as coarse-grainable, intermittent compaction. Then, and only then, does  {\sc gsh} provide an 
understanding  for tapping -- though this will be a transparent, conventional and demystified one.

\section{Shear Bands\label{sb}} 

A shear band is in its essence the coexistence of static granular solid and uniform dense flow. In 
the first, the grains are deformed and at rest, $T_g=0$, with all energy being elastic. In the 
second, the grains jiggle, rattle, move macroscopic distances,  with $T_g\sim v_s$ and a  portion 
of the energy in $T_g$. 

The transition from the rate-independent critical state to the Bagnold 
regime of uniform dense flow as the shear rate  $v_s$ increases go via two different paths, either gradual and uniform, as discussed in Sec~\ref{udf};
or discontinuous and nonuniform, via shear bands. 

Approaching the critical state with a high initial density, the evolution of the shear stress 
$\sigma_s$ is non-monotonic, assuming values temporarily larger than $\sigma_c$. This is 
where the system has a high probability of breaching an instability, either of the elastic energy at a point on the yield surface, as discussed in Sec~\ref{yield surfaces}, or that of $T_g$, as discussed in Sec~\ref{scs}. The breaching of the elastic energy will happen 
with certainty if the system is slowly sheared in the quasi-elastic regime. After the breach, the density and elastic strain quickly become inhomogeneous, because their fluctuations 
grow exponentially. This goes on until a stable state 
compatible with the boundary conditions is found again -- such as one with a shear band, 
consisting of a low-density fluid region in the shear band, and a high density solid region 
outside. The chaotic transition is difficult to account for, but the stable shear band is again 
simple.

As we shall see, the shear band has a minimal and constant width at a low shearing velocity $v$. If $v$ is higher, the system's behavior depends on the setup. For given pressure, the width $\ell$ grows linearly with $v$, implying a constant rate $v/\ell$ in the liquid phase. As a result, the shear stress, a function of the rate, remains independent of $v$. This {\em faux rate-independence} goes on until the band covers the whole system, at which point the quadratic rate dependence of uniform dense flow sets in. 
For given volume, the band width remains independent of $v$, but the shear stress grows quadratically with it. The transition to uniform dense flow is again discontinuous. It happens when the shear stress exceeds the critical value of the solid density, at which point the solid phase is no longer stable.

To account for the shear band, we connect the fluid and solid solutions already considered employing a set of simple connecting conditions. Denoting the solid and fluid parts  respectively with the superscripts $^S$ and $^F$, the conditions are the equality of the pressure, shear stress, and chemical potential,
\begin{equation}\label{3b-21}
P^S=P^F,\quad \sigma_s^S=\sigma_s^F, \quad\mu^S=\mu^F.
\end{equation}
[The chemical potential is defined as $\mu\equiv\partial w/\partial\rho$, Eq~(\ref{2-2}). The equality holds because otherwise a particle current would flow across the phase boundary.]  All three fields have an elastic and a seismic contribution, Eqs~(\ref{2c-2},\ref{2c-2a}). With $P=P_\Delta+P_T$, $\mu=\mu_\Delta+\mu_T$, they are
\begin{eqnarray}\label{5-7a}
P_\Delta&\equiv&({1-\alpha})\Delta^{1.5}\left[{\cal B} +
{\cal A}{u_s^2}/({2\Delta^2})\right]
\\\label{5-7d}
P_T&\equiv&
{a\rho T_g^2}/2({\rho_c/\rho-1}),
\\\label{5-7}
\mu_T&\equiv&T_g^2\,\frac b2\frac{(1+a)\rho-\rho_{cp}}{\rho_{cp}-\rho},
\\\mu_\Delta&\equiv&\frac{0.15w_\Delta
}{\rho_{cp}-\rho}\,\,\frac{\rho_{cp}-\bar\rho}
{\rho-\bar\rho},\end{eqnarray}
in addition to ${\sigma_s}=2({1-\alpha}){\cal A}\,u_s\sqrt\Delta-\eta_1T_g v_s$, also with two parts.  Denoting  the width of the shear band as $\ell$, and the velocity difference across the shear band as $v$,  we assume
\begin{eqnarray}
\text{in fluid:}&\quad& v_s=v/\ell\sim T_g,\,\, \Delta^F=\Delta_c,\,\, u_s^F=u_c,
\\
\text{in solid:}&\quad& \alpha, T_g, v_s=0.
\end{eqnarray}
In other words, the elastic strain $\Delta$ and $u_s$ have critical values in the $F$-phase, and appropriate static values in the $S$-phase.  Strictly speaking, the discontinuities at the $S-F$ boundary are in $\rho,\Delta, u_s$, but not the shear rate, which decays exponentially in $S$, as a result of $T_g$-diffusion, see Sec~\ref{creep motion}. We neglect this detail in the qualitative discussion below.

\subsection{The Fluid Region}
The elastic contribution $\mu_\Delta$ is a very small quantity: In $P_\Delta\sim{\cal B}\Delta^{1.5}$, a large ${\cal B}$ compensates a small $\Delta^{1.5}$, such that $P_\Delta$ is either much larger than, or comparable to, $P_T\sim T_g^2$. Now,  $\mu_T$ is of the order of $P_T/\rho$, but $\mu_\Delta\sim{\cal B}\Delta^{2.5}\sim \Delta P_\Delta$ is smaller by the factor $\Delta$, around $10^{-3}-10^{-4}$. Therefore, as long as $P_T\gg\Delta P_\Delta$, we have  $\mu_T\gg\mu_\Delta$, and $\mu^S=\mu^F$ reduces to $\mu_T=0$, implying  the density in the shear band is fixed,
\begin{equation}\label{5-8}
\rho^{F}=\rho_{cp}/(1+a).
\end{equation}
Measuring this density therefore yields the value of $a$ [that is important in calibrating the energy contribution of granular entropy, see Eq~(\ref{2b-5})]. Note that given $\rho^F$, the elastic pressure $P_\Delta$ is also known, because $\Delta^F=\Delta_c (\rho^F), u_s^F=u_c(\rho^F)$. 

\subsubsection{Given Pressure}
\begin{figure}[t]
\begin{center}
\includegraphics[scale=.6]{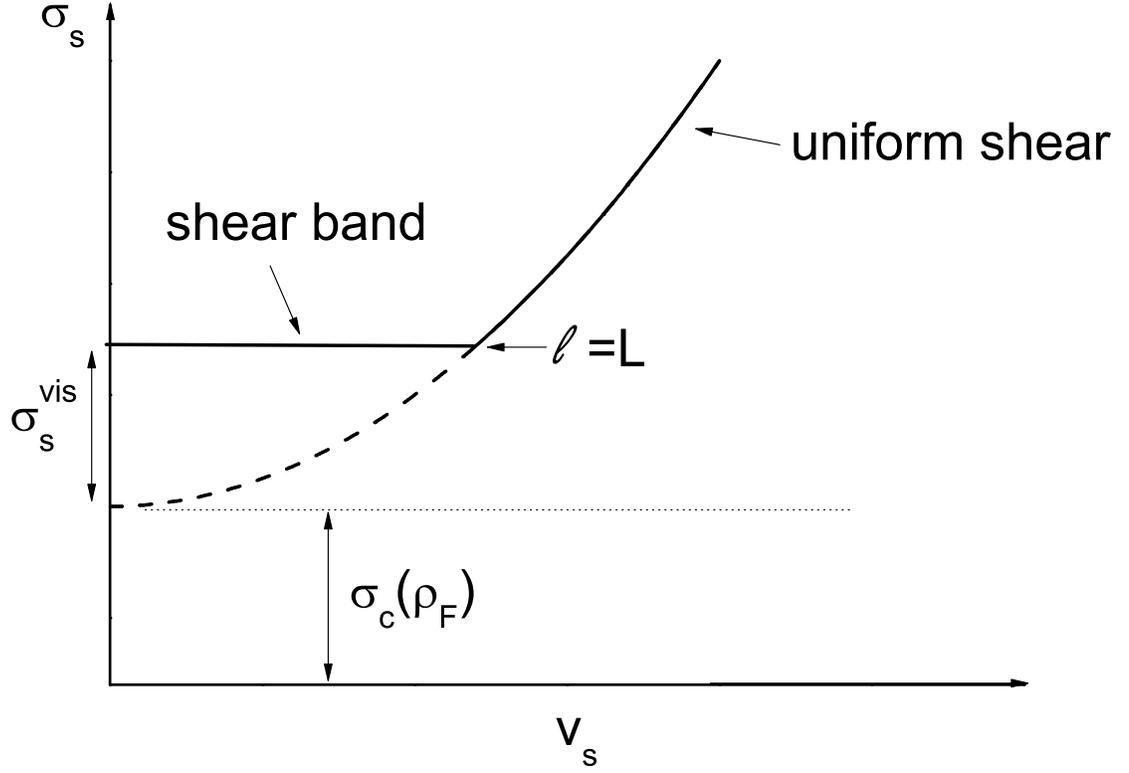}
\end{center}\caption{Shear stress $\sigma_s$ as functions of the velocity difference $v$  for given pressure, in a simple-shear geometry. The offset gives the elastic contribution, $\sigma_c(\rho_F)$; the parabola is the case without a shear band. The thick horizontal line depicts the situation with a shear band, of width $\ell$, which is smaller towards left, and equal to the system's width $L$ at the right end. The rate-independence of $\sigma_s$ derived from  $\ell$ adjusting itself such that $T_g\sim v/\ell$ remains constant for given pressure.  \label{fig4}} 
\end{figure}

Next, we confine the discussion to the case of given external pressure, $P^{ex}=P^S=P^F$ and given velocity difference $v$ across the shear band.
This is an intriguing case, because $P^F$ and $\rho^F$ fix both $T_g$ and the shear rate 
$T_g\sim v_s=v/\ell=T_g\sqrt{\gamma_1/\eta_1}$. Given in addition $v$, the width $\ell$ of the 
shear band is also fixed. These are all there is to be known about the fluid region. Especially 
the pressure and the shear stress are given as
\begin{eqnarray}
P&=&P_c(\rho^F)+\frac{T_g^2}{2}\,\frac{(\rho^F)^2\,a\,b/\rho_{cp}}{(1-\rho^F/\rho_{cp})},
\\
\sigma_s&=&\sigma_c(\rho^F) -\eta_1T_g\,v/\ell.
\end{eqnarray}
Remarkably, the system now displays a faux rate-independence: $\ell$ adjusts itself such that $T_g\sim v/\ell$ remains constant for given pressure,
independent what $v$ is. The parabola of Fig~\ref{fig4} depicts  $\sigma_s$. The offset gives the elastic contributions, $\sigma_c$. The horizontal line is a result of  $\ell$ adjusting.   
It is indeed easy to mistake a shear band for the uniform, critical state.

Increasing the velocity $v$ at given pressure alters the width $\ell$, as long as it is smaller than the width of the total system $L$. For larger velocities, the system is again uniform, without a solid region. And the consideration of Sec~\ref{udf} holds. Until this point, the stress is rate-independent, much longer than without a shear band. 

Given the solid density $\rho^S$ (which is fixed by the dynamics, see Sec~\ref{sr}) and the mass per unit length $M$, mass conservation $\rho^S(L-\ell)+\rho^F\ell=M$ determines the total width $L$ for given pressure $P$. 

\subsubsection{Given Total Volume}
At given total volume $L$, the behavior is quite different. First, because of mass conservation,
\begin{equation}
\rho^S(L-\ell)+\rho^F\ell=M,
\end{equation}
and because $\rho^S,\rho^F$ are given in addition to $L$, the band width $\ell$ is fixed, irrespective what the velocity $v$ is. As a result, both the shear stress and pressure grow as $(v/\ell)^2\sim v^2$, not at all rate-independent. The transition to uniform dense flow happens when Eq~(\ref{94}) is violated, for $\sigma_c(\rho^S)=\sigma^S=\sigma_c(\rho^F)+\eta_1T_gv_s$.

\subsection{The Solid Region\label{sr}}
The  solid region is, in comparison, less fixed. The reason is we have the three connecting conditions of Eq~(\ref{3b-21}), and three quantities to be determined,  $\Delta^S, u_s^S, \rho^S$. Yet,  
because terms are of such different magnitudes in $\mu^S=\mu^F$, it fixes $\rho^F$ instead of 
giving a relation between $\rho^F$ and  $\rho^S$. So it is always satisfied, irrespective what 
value $\rho^S$ assumes. Therefore, $\rho^S$ can only be  a result of the dynamics: When an instability is breached, the density is changed until it gets stuck at some value for 
$\rho^S$, at which the system is again stable. Then of course, $\Delta^S, u_s^S$ may be determined for given pressure and shear stress.
Nevertheless, we do know  
\begin{equation}\label{92}
\rho^F<\rho^S \quad\text{and}\quad \rho^F\le\rho_c
\end{equation}
must hold. The first inequality can be seen from 
\begin{equation}\label{94}
\sigma_c(\rho^S)>\sigma^S=\sigma_c(\rho^F)+\eta_1T_gv_s\ge\sigma_c(\rho^F).
\end{equation}
The first greater sign is related to the discussion in Sec~\ref{scs}; the equal sign is one connecting condition; and the second greater sign is a result of $\eta_1T_gv_s$ being positive, in addition to the fact that $\sigma_c$ is a monotonically increasing function of the density, cf. the discussion below Eq~(\ref{3b-5}). The second inequality, $\rho^F\le\rho_c$, holds for given external pressure $P$, and comes from the following consideration: 
In the critical state, there is only one free parameter. Once $\rho$ is given, $\Delta_c,u_c,P_c,
\sigma_c$ also are. Alternatively, one may fix the external pressure $P=P_c\equiv (1-\bar\alpha)
P^c_\Delta(\rho_c)$, then $\rho_c(P)$ is a dependent quantity. In the shear band, because the 
density $\rho^F$ is already fixed, the elastic pressure $P_c(\rho^F)$ will in general be different 
from the external one, $P=P_c(\rho_c)$, and the difference is taken up by the seismic term 
$P_T$. Since  $P_T$ is always positive, we have $P_c(\rho^F)<P_c( \rho_c)$, implying the 
external pressure has to be so large that $\rho^F\le\rho_c$ holds. Otherwise,  shear band cannot exist, and the flow is uniform.

\subsection{Minimal Band Width\label{minimal band width}}
When the velocity $v$ decreases, the above consideration stops to be valid at some point. For instance, $\rho^F$ is no longer given if $P_T\gg\Delta P_\Delta$ does not hold. More importantly, the width $\ell$ will decrease with $v$ (for given pressure) only as long as $\ell$ is larger than a few grain diameter. When $v$ decreases further, $\ell$ cannot follow, and will remain at a minimal width. To account for this, we need the consideration (as yet quite qualitative) given below.  

The phenomenon of clogging implies that a free surface, if small enough, may be stable even when facing downward, with a friction angle of up to $180^\circ$, 
much larger than the coulomb yield angle.  Similarly, shear bands have a a finite width in the limit of vanishing velocities. Both are phenomena not accounted for by {\sc gsh} as given above. This is connected to the fact that hydrodynamic theories  are only capable of accounting for spatial variations much larger than the correlation length -- in the case of {\sc gsh} especially the grain diameter. Nevertheless, there is a tried and proven method of qualitatively accounting for small scale effects such that blatant inconsistencies are avoided. (It has been employed eg. for the superfluid  transition by including the gradient terms of the order parameter's magnitude~\cite{Khal}.)

In our case, we include higher-order gradient terms $\sim(\nabla_k u_{ij})^2$ in the energy that express the extra cost of a nonuniform strain field. A length scale on which elastic strains will 
change is thus introduced. Note non-uniform strain fields necessarily exist at the liquid-solid interface, and an infinitely narrow shear band is the result of setting 
the length of strain change to zero. Similarly, a non-uniform strain field of the size of the hole's diameter is needed for unclogging. 

Because of momentum conservation, $\nabla_j\sigma_{ij}=0$, the stress stays constant in one-dimensional geometries, even across a shear shear band. Therefore, higher order gradient terms in the stress would not do the job. Including higher order gradient terms in strain rates would also miss the point that a static inhomogeneity of granular deformation costs extra energy and is therefore capped~\cite{wu2}. 
The lack of a length scale in describing shear bands is a known problem in 
soil mechanics. One popular method to introduce it is by adding state variables that account for 
the couple stress and the Crosserat rotation, see eg.~\cite{wu1}. Including additional variables for the sole purpose of solving our present problem, however, does seem unwarranted as it  leads to a far more complex theory.   

Starting with an addition to the elastic energy $\sim(\nabla_ku_{ij})^2$ 
and introducing the conjugate variable $\phi_{ijk}\equiv\partial w/\partial \nabla_ku_{ij}$, the 
elastic stress obtains additional terms of the type
\begin{equation}\label{GSE3}
\nabla_k\phi_{ijk}\sim \nabla^2u_{ij}.
\end{equation}
In a shear band of width $\ell$, we therefore expect an additional pressure contribution $P_{\ell}\sim1/\ell^2$, which is to be compared with $P_T\sim T_g^2\sim v^2/\ell^2$. 
Defining $A$ such that $P_T=(Av/\ell)^2$, and $v_0$ such that $P_\ell= (Av_0/\ell)^2$, where $v_0$ is a function of the elastic strain and its difference at the interface, the total pressure $P=P^F=P^S$ is
\begin{equation}
P=P_c(\rho^F)+(A/\ell)^2(v^2+v_0^2).
\end{equation}
As long as $v$ is fast enough for $\ell$ to be larger than, say, 30 grain diameter, $v_0\ll v$ may be neglected, and the results of the last sections is recovered. But in the vicinity of a few grain diameter, it becomes dominant, and fixes the band width to a value independent of $v$. As shear bands are usually observed to be narrow and rate independent, experiments are probably typically in this limit.  
A constant pressure contribution $\sim1/\ell^2$ will also stabilize a free surface of diameter $\ell$ that is sufficiently small -- a subject that we shall consider elsewhere, along with a  more quantitative consideration of shear bands.  

%
%
%


\begin{thebibliography}{99} 

\bibitem{schofield}
P.~Wroth A.~Schofield.
\newblock {\em Critical State Soil Mechanics}.
\newblock McGraw-Hill, London, 1968.

\bibitem{nedderman}
R.M. Nedderman.
\newblock {\em Statics and Kinematics of Granular Materials}.
\newblock Cambridge University Press, 1992.

\bibitem{wood1990}
D.~M. Wood.
\newblock {\em Soil Behaviour and Critical State Soil Mechanics}.
\newblock Cambridge University Press, 1990.

\bibitem{kolymbas1}
D.~Kolymbas.
\newblock {\em Introduction to Hypoplasticity}.
\newblock Balkema, Rotterdam, 2000.

\bibitem{kolymbas2}
W.~Wu and D.~Kolymbas.
\newblock {\em Constitutive Modelling of Granular Materials}.
\newblock Springer, Berlin, 2000.

\bibitem{gudehus2010}
G.~Gudehus.
\newblock {\em Physical Soil Mechanics}.
\newblock Springer SPIN, 2010.

\bibitem{hutter2007}
S.P. Pudasaini and K.~Hutter.
\newblock {\em Avalanche Dynamics}.
\newblock Springer, 2007.

\bibitem{LL6}
L.~D. Landau and E.~M. Lifshitz.
\newblock {\em Fluid Mechanics}.
\newblock Butterworth-Heinemann, 1987.

\bibitem{Khal}
I.~M. Khalatnikov.
\newblock {\em Introduction to the Theory of Superfluidity}.
\newblock Benjamin, New York, 1965.

\bibitem{deGennes}
P.G. de~Gennes and J.~Prost.
\newblock {\em The Physics of Liquid Crystals}.
\newblock Clarendon Press, Oxford, 1993.

\bibitem{hydro-1}S. R. de Groot and P. Masur,
{\it Non-Equilibrium Thermodynamics}, (Dover, New York 1984).

\bibitem{hydro-2}D. Forster, Hydrodynamic Fluctuations, Broken Symmetry and
Correlation Functions (Benjamin, New York, 1975).

\bibitem{liqCryst-1} P.G. de Gennes and J. Prost, {\em The Physics of
Liquid Crystals} (Clarendon Press, Oxford 1993). 

\bibitem{liqCryst-2} P.C.
Martin, O. Parodi, and P.S. Pershan, {\it Unified Hydrodynamic Theory for
Crystals, Liquid Crystals, and Normal Fluids}, Phys. Rev. A 6, 2401 (1972).

\bibitem{liqCryst-3} T.C. Lubensky, {\it Hydrodynamics of Cholesteric
Liquid Crystals}, Phys. Rev. A 6, 452 (1972). 

\bibitem{liqCryst-4} M. Liu,
{\it Hydrodynamic Theory near the Nematic Smectic-A Transition}, Phys. Rev.
{\bf A 19}, 2090 (1979); 

\bibitem{liqCryst-5} M. Liu, {\it Hydrodynamic
theory of biaxial nematics}, Phys. Rev. {\bf A 24}, 2720 (1981).
\bibitem{liqCryst-6} M. Liu, {\it Maxwell equations in nematic liquid
crystals}, Phys. Rev. {\bf E 50}, 2925, (1994). 

\bibitem{liqCryst-7} H.
Pleiner and H.R. Brand, in {\it Pattern Formation in Liquid Crystals},
edited by A. Buka and L. Kramer (Springer, New York, 1996).

\bibitem{he3-1} R. Graham, {\it Hydrodynamics of 3He in Anisotropic A
Phase}, Phys. Rev. Lett. {\bf 33}, 1431 (1974). 

\bibitem{he3-2} R. Graham
and H. Pleiner, {\it Spin Hydrodynamics of 3He in the Anisotropic A Phase},
Phys. Rev. Lett. {\bf 34}, 792 (1975). 

\bibitem{he3-3} M. Liu, {\it
Hydrodynamics of $^3$He near the A-Transition,} Phys. Rev. Lett. {\bf 35},
1577 (1975). 

\bibitem{he3-4} M. Liu and M.C. Cross, {\it Broken Spin-Orbit
Symmetry in Superfluid $^3$He and the B-Phase Dynamics,} Phys. Rev. Lett.
{\bf 41}, 250 (1978). 

\bibitem{he3-5} M. Liu and M.C. Cross, {\it Gauge
Wheel of Superfluid $^3$He,} Phys. Rev. Lett. {\bf 43}, 296 (1979).

\bibitem{he3-6} M. Liu, {\it Relative Broken Symmetry and the Dynamics of
the $A_1$-Phase,} Phys. Rev. Lett. {\bf 43}, 1740 (1979).

\bibitem{SC-1} M. Liu, {\it Rotating Superconductors and the
Frame-independent London Equations,} Phys. Rev. Lett. {\bf 81}, 3223,
(1998). 

\bibitem{SC-2} Jiang Y.M. and M. Liu, {\it Rotating Superconductors
and the London Moment: Thermodynamics versus Microscopics,} Phys. Rev. {\bf
B 6}, 184506, (2001). 

\bibitem{SC-3} M.~Liu, {\em Superconducting
Hydrodynamics and the Higgs Analogy,} J. Low Temp. Phys. 126, 911, (2002)

\bibitem{hymax-1} K. Henjes and M. Liu, {\it Hydrodynamics of Polarizable
Liquids,} Ann. Phys. {\bf 223}, 243 (1993). 

\bibitem{hymax-2} M. Liu, {\it
Hydrodynamic Theory of Electromagnetic Fields in Continuous Media,} Phys.
Rev. Lett. {\bf 70}, 3580 (1993). 

\bibitem{hymax-3} {\it Mario Liu
replies,} Phys. Rev. Lett. {\bf 74}, 1884, (1995). 

\bibitem{hymax-4} Y.M.
Jiang and M. Liu, {\it Dynamics of Dispersive and Nonlinear Media,} Phys.
Rev. Lett. {\bf 77}, 1043, (1996).

\bibitem{FF-1} M.I. Shliomis, {\em Magnetic Fluids}, Sov. Phys. Usp. 17,
153 (1974). 

\bibitem{FF-2} R.E. Rosensweig, {\em Ferrohydrodynamics},
(Dover, New York 1997). 

\bibitem{FF-3} M. Liu, {\it Fluiddynamics of
Colloidal Magnetic and Electric Liquid,} Phys. Rev. Lett. {\bf 74}, 4535
(1995). 

\bibitem{FF-4} M. Liu, {\it Off-Equilibrium, Static Fields in
Dielectric Ferrofluids,} Phys. Rev. Lett. {\bf 80}, 2937, (1998).

\bibitem{FF-5} M. Liu, {\it Electromagnetic Fields in Ferrofluids}, Phys.
Rev. {\bf E 59}, 3669, (1999). 

\bibitem{FF-6} H.W.~M\"{u}ller and M.~Liu,
{\it Structure of Ferro-Fluiddynamics,} Phys. Rev. {\bf E 64}, 061405
(2001). 

\bibitem{FF-7} H.W. M\"{u}ller and M. Liu, {\em Shear Excited Sound
in Magnetic Fluid}, Phys. Rev. Lett. {\bf 89}, 67201, (2002).

\bibitem{FF-8} O. M\"{u}ller, D. Hahn and M. Liu, {\em Non-Newtonian
behaviour in ferrofluids and magnetization relaxation,} J. Phys.: Condens.
Matter 18, 2623, (2006). 

\bibitem{FF-9} S. Mahle, P. Ilg and M. Liu, {\em
Hydrodynamic theory of polydisperse chain-forming ferrofluids,} Phys. Rev.
{\bf E 77}, 016305 (2008).

\bibitem{polymer-1} H. Temmen, H. Pleiner, M. Liu and H.R. Brand, {\it
Convective Nonlinearity in Non-Newtonian Fluids,} Phys. Rev. Lett. {\bf
84}, 3228 (2000). 

\bibitem{polymer-2}H. Temmen, H. Pleiner, M. Liu and H.R.
Brand,{\it Temmen et al. reply}, Phys. Rev. Lett. {\bf 86}, 745 (2001).

\bibitem{polymer-3} H. Pleiner, M. Liu and H.R. Brand, {\it Nonlinear Fluid
Dynamics Description of non-Newtonian Fluids}, {Rheologica Acta} {\bf 43},
502 (2004). 

\bibitem{polymer-4}O. M\"{u}ller, {\em Die Hydrodynamische
Theorie Polymerer Fluide}, PhD Thesis University T\"{u}bingen (2006).

\bibitem{midi}
GDR MiDi.
\newblock On dense granular flows.
\newblock {\em The European Physical Journal E}, 14(4):341--365 (2004).

\bibitem{kadanoff}
L.~P. Kadanoff.
\newblock Built upon sand: Theoretical ideas inspired by granular flows.
\newblock {\em Reviews of Modern Physics}, 71 (1):435 -- 444 (1999).

\bibitem{Houlsby}
G.~T. Houlsby and A.~M. Puzrin.
\newblock {\em Principles of Hyperplasticity}.
\newblock Springer (2006).

\bibitem{Houlsby2}
I.~F. Collins and G.~T. Houlsby.
\newblock Application of thermomechanical principles to the modelling of
  geotechnical materials.
\newblock {\em Proc. R. Soc. Lond. A}, 453:1975--2001, 1997.


\bibitem{granR2}
Y.~Jiang and M.~Liu.
\newblock Granular solid hydrodynamics.
\newblock {\em Granular Matter}, 11:139, May 2009.
Free download: http://www.springerlink.com/content/a8016874j8868u8r/fulltext.pdf

\bibitem{granR3}
Y.~Jiang and M.~Liu.
\newblock The physics of granular mechanics.
\newblock In D.~Kolymbas and G.~Viggiani, editors, {\em Mechanics of Natural
  Solids}, pages 27--46. Springer, 2009.


\bibitem{gudehus-jl}
G.~Gudehus, Y.M. Jiang, and M.~Liu.
\newblock Seismo- and thermodynnamics of granular solids.
\newblock {\em Granular Matter}, 1304:319--340, 2011.

\bibitem{luding2009}
Stefan Luding.
\newblock Towards dense, realistic granular media in 2d.
\newblock {\em Nonlinearity}, 22:101--146, 2009.

\bibitem{Bocquet}
L.~Bocquet, W.~Losert, D.~Schalk, T.~C. Lubensky, and J.~P. Gollub.
\newblock Granular shear flow dynamics and forces: Experiment and continuum
  theory.
\newblock {\em Phys. Rev. E}, 65(1):011307, Dec 2001.

\bibitem{denseFlow}
Stefan~Mahle, Yimin~Jiang and Mario~Liu.
\newblock Granular solid hydrodynamics: Dense flow, fluidization and jamming.
\newblock {\em arXiv:1010.5350v1 [cond-mat.soft]}, 2010.


\bibitem{ge1}
D.~O. Krimer, M.~Pfitzner, K.~Br\''auer, Y.~Jiang, and M.~Liu.
\newblock Granular elasticity: General considerations and the stress dip in
  sand piles.
\newblock {\em Phys. Rev. E)}, 74(6):061310, 2006.

\bibitem{ge2}
K.~Br\"auer, M.~Pfitzner, D.~O. Krimer, M.~Mayer, Y.~Jiang, and M.~Liu.
\newblock Granular elasticity: Stress distributions in silos and under point
  loads.
\newblock {\em Phys. Rev. E (Statistical, Nonlinear, and Soft Matter Physics)},
  74(6):061311, 2006.

\bibitem{kuwano2002}
R.~Kuwano and R.~J. Jardine.
\newblock On the applicability of cross-anisotropic elasticity to granular
  materials at very small strains.
\newblock {\em Geotechnique}, 52(10):727--749, Dec 2002.

\bibitem{ge3}
Y.~Jiang and M.~Liu.
\newblock Incremental stress-strain relation from granular elasticity:
  Comparison to experiments.
\newblock {\em Phys. Rev. E (Statistical, Nonlinear, and Soft Matter Physics)},
  77(2):021306, 2008.
  
\bibitem{jia2009}
Y.~Khidas and X.~Jia.
\newblock Anisotropic nonlinear elasticity in a spherical-bead pack: Influence
  of the fabric anisotropy.
\newblock {\em Phys. Rev. E}, 81:021303, Feb. 2010.


\bibitem{ge4}
M.~Mayer and M.~Liu.
\newblock Propagation of elastic waves in granular solid hydrodynamics.
\newblock {\em Phys. Rev. E}, 82:042301, 2010.

\bibitem{hardin}
B.O. Hardin and F.E. Richart.
\newblock Elastic wave velocities in granular soils.
\newblock {\em J. Soil Mech. Found. Div. ASCE}, 89: SM1:33--65, 1963.

\bibitem{lade-duncan}
P.V. Lade and J.M. Duncan.
\newblock Elastoplastic stress-strain theory for cohesionless soil.
\newblock {\em Proc. ASCE, JGTD,}, 101:N0 GT10, 1975.

\bibitem{matsuoka}
H.~Matsuoka and T.~Nakai.
\newblock Stress-strain relationship of soil based on the smp.
\newblock {\em Proc. 9th ICSMFE}, specialty session 9:153--163, 1977.

\bibitem{3inv} Y.M. Jiang, H.P. Zheng, Z. Peng, L.P. Fu, S.X. Song, Q.C. Sun, M. Mayer, and M. Liu, \newblock Expression for the granular elastic energy.
\newblock  Phys. Rev. E {\bf 85}, 051304 (2012)    


\bibitem{granL3}
Y.~Jiang and M.~Liu.
\newblock From elasticity to hypoplasticity: Dynamics of granular solids.
\newblock {\em Phys. Rev. Lett.}, 99(10):105501, 2007.
\bibitem{roux} J.-N. Roux.  How granular materials deform in quasistatic conditions  AIP Conf. Proc. 1227, pp. 260-270; doi:http://dx.doi.org/10.1063/1.3435396; The nature of quasi-static deformation in granular materials. {\em arXiv:0901.2305v1 [cond-mat.soft]}, 2009;
\bibitem{critState}
Stefan Mahle, Yimin Jiang, and Mario Liu.
\newblock The critical state and the steady-state solution in granular solid
  hydrodynamics.
\newblock {\em arXiv:1006.5131v3 [physics.geo-ph]}, 2010.


\bibitem{vHecke2011}
Joshua~A. Dijksman, Geert~H. Wortel, Olivier van Dellen, Loevrens T.H.~Dauchot,
  and Martin van Hecke.
\newblock Jamming, yielding, and rheology of weakly vibrated granular media.
\newblock {\em PRL}, page 108303, 2011.

\bibitem{aging}
Van~Bau Nguyen, Thierry Darnige, Ary Bruand, and Eric Clement.
\newblock Creep and fluidity of a real granular packing near jamming.
\newblock {\em Phys. Rev. Lett}, 107:138303, 2011.

\bibitem{komatsu}
T.S. Komatsu, S.~Inagaki, N.~Nakagawa, and S.~Nasuno.
\newblock Creep motion in a granular pile exhibiting steady surface flow.
\newblock {\em Phys. Rev. Lett.}, 86:1757�1760, 2001.

\bibitem{crassous}
J~Crassous, J-F Metayer, P~Richard, and C.~Laroche.
\newblock Experimental study of a creeping granular flow at very low velocity.
\newblock {\em J. Stat. Mech.}, 2008:P03009, 2008.


\bibitem{nichol2010}
Kiri Nichol, Alexey Zanin, Renaud Bastien, Elie Wandersman, and Martin van
  Hecke.
\newblock Flow-induced agitation creates a granular fluid.
\newblock {\em Phys. Rev. Lett.}, 104:078302, 2010.

\bibitem{reddy2011}
K.A. Reddy, Y.~Forterre, and O.~Pouliquen.
\newblock Evidence of mechanical activated processes in slow granular flows.
\newblock {\em Phys. Rev. Lett.}, 106:108301, 2011.

\bibitem{jia2004}
X.~Jia.
\newblock Codalike multiple scattering of elastic waves in dense granular
  media.
\newblock {\em Phys. Rev. Lett.}, 93(15):154303, Oct 2004.

\bibitem{jia1999}
X.~Jia, C.~Caroli, and B.~Velicky.
\newblock Ultrasound propagation in externally stressed granular media.
\newblock {\em Phys. Rev. Lett.}, 82(9):1863--1866, Mar 1999.

\bibitem{zhang2012}
Q.~Zhang, Y.C. Li, M.Y. Hou, Y.M. Jiang, and M.~Liu.
\newblock Elastic waves in the presence of a granular shear band formed by
  direct shear.
\newblock {\em Phys. Rev. E}, 85:031306, 2012.



\bibitem{barodesy} Kolymbas D. Barodesy: a new constitutive frame for soils. Geotechnique Letters 2, 17–23,  (2012), http://dx.doi.org/10.1680/geolett.12.00004; 
Barodesy: A new hypoplastic approach. International Journal for Numerical and Analytical Methods in Geomechanics (2011). doi:10.1002/nag.1051;
Sand as an archetypical natural solid. In Mechanics of Natural Solids, Kolymbas D, Viggiani G (eds.). Springer: Berlin, (2009); 1–26; 

\bibitem{GSH&Barodesy} Yimin Jiang, and Mario Liu. Proportional Path, Barodesy, and Granular Solid Hydrodynamics. Preprint


\bibitem{Bagnold}
R.~A. Bagnold.
\newblock Experiments on a gravity-free dispersion of large solid spheres in a
  {N}ewtonian fluid under shear.
\newblock {\em Proceedings of the Royal Society of London. Series A.
  Mathematical and Physical Sciences}, 225(1160):49--63, 1954.
  
  
\bibitem{pouliquen1}
Pierre Jop, Yo\"{e}l Forterre, and Olivier Pouliquen.
\newblock A constitutive law for dense granular flows.
\newblock {\em Nature}, 441:727--730, 2006.

\bibitem{pouliquen2}
Yo\"{e}l Forterre and Olivier Pouliquen.
\newblock Flows of dense granular media.
\newblock {\em Annu. Rev.Fluid Mech.}, 40:1--24, 2008.


\bibitem{Savage}
S.~B. Savage and M.~Sayed.
\newblock Stresses developed by dry cohesionless granular materials sheared in
  an annular shear cell.
\newblock {\em Journal of Fluid Mechanics Digital Archive}, 142:391--430, 1984.

\bibitem{aranson}
I.~S. Aranson and L.~S. Tsimring.
\newblock Continuum theory of partially fluidized granular flows.
\newblock {\em Phys. Rev. E}, 65:061303,, 2002.

\bibitem{Brodsky2}
K.~Lu, E.E. Brodsky, and H.P. Kavehpour.
\newblock {\em J. Fluid. Mech.}, 587:347, 2007.

\bibitem{Brodsky1}
K.~Lu, E.E. Brodsky, and H.P. Kavehpour.
\newblock {\em Nature Letters}, 4:404, 2008.

\bibitem{dijksman2011}J.A. Dijksman, G.H. Wortel, L.T.H. van Dellen,
O. Dauchot, and M. van Hecke, Phys. Rev. Lett. {\bf 107},
108303(2011).

\bibitem{mem}C. Josserand, A.V. Tkachenko, D.M.
Mueth, H.M. Jaeger, Phys. Rev. Lett., \textbf{85}, 3632
(2000)


\bibitem{1nico} P. Richard, M. Nicodemi, R. Delannay,
P. Ribiere, D. Bideau, Nature, {\bf 4}, 121 (2005)


\bibitem{Edw} S.F. Edwards, R.B.S. Oakeshott,
Physica A{\bf 157}, 1080 (1989); S.F. Edwards, D.V.
Grinev, Granular Matter,  {\bf 4}, 147 (2003).

\bibitem{compaction} Yimin Jiang, and Mario Liu.
\newblock The critical state and the steady-state solution in granular solid
  hydrodynamics.
\newblock {\em arXiv:0911.2199v2 [cond-mat.soft]}, 2010.
\bibitem{wu2}Wei Wu. {\em On high-order hypoplastic models for granular materials}. Journal of Engineering Mathematics {\bf 56}: 23–34 (2006) 

\bibitem{wu1}Tejchman, J. and Wu, W. {\it FE-investigations of micro-polar boundary conditions along interface between soil and structure}, Granular Matter, {\bf 12}, 399 (2010)


\end{thebibliography}

  \end{document}